\documentclass[prd,notitlepage,aps,a4paper,11pt,amsmath,amsfonts,amssymb,nofootinbib,superscriptaddress]{revtex4-1}
\usepackage{enumerate}
\usepackage[T1]{fontenc}
\usepackage[utf8x]{inputenc}
\usepackage[colorlinks=true]{hyperref}

\usepackage{mathtools}

\begin{document}

\title{Geodesics and the magnitude-redshift relation on cosmologically symmetric Finsler spacetimes}

\author{Manuel Hohmann}
\email{manuel.hohmann@ut.ee}
\affiliation{Laboratory of Theoretical Physics, Institute of Physics, University of Tartu, W. Ostwaldi 1, 50411 Tartu, Estonia}

\author{Christian Pfeifer}
\email{christian.pfeifer@ut.ee}
\affiliation{Laboratory of Theoretical Physics, Institute of Physics, University of Tartu, W. Ostwaldi 1, 50411 Tartu, Estonia}
\affiliation{Institute for Theoretical Physics, University of Hannover, Appelstrasse 2, 30167 Hannover, Germany}
\affiliation{Center of Applied Space Technology and Microgravity (ZARM), University of Bremen, Am Fallturm, 28359 Bremen, Germany}

\begin{abstract}
We discuss the geodesic motion of both massive test particles, following timelike geodesics, and light, following null geodesics, on Finsler spacetimes with cosmological symmetry. Using adapted coordinates on the tangent bundle of the spacetime manifold, we derive the general form of the geodesic equation. Further, we derive a complete set of constants of motion. As an application of these findings, we derive the magnitude-redshift relation for light propagating on a cosmologically symmetric Finsler background, both for a general Finsler spacetime and for particular examples, such as spacetimes equipped with Bogoslovsky and Randers length measures. Our results allow a confrontation of these geometries with observations of the magnitude and redshift of supernovae.
\end{abstract}

\maketitle

%%%%%%%%%%%%%%%%%%%%%%%%%%%%%%%%%%%%%%%%%%%%%%%%%%%%%%%%%%%%%%%%%%%%%%%%%%%%%%%%%%
\section{Introduction}
One of the most revolutionary observations in modern cosmology is the measurement of the magnitude-redshift relation of distant supernovae~\cite{Riess:1998cb,Perlmutter:1998np,Kowalski:2008ez,Amanullah:2010vv,Suzuki:2011hu}. It follows from the kinematics of a homogeneous and isotropic universe, whose geometry is modeled by a Friedmann-Lemaitre-Robertson-Walker (FLRW) metric, that this relation allows to directly measure the deceleration or acceleration of the expansion of the universe in terms of a single parameter \(q\), called the deceleration parameter. Numerous analyses of supernova data have come to results in the range \(-1.0 < q < -0.5\) at the present epoch~\cite{John:2004vf,Gong:2006gs,Cunha:2008mt,Li:2010da,Giostri:2012ek,Mukherjee:2016trt}, where a negative value indicates an accelerating expansion. This result clearly contradicts the expected behavior of a universe described by general relativity and filled with perfect fluid matter with non-negative barotropic index \(w \geq 0\), whose expansion should decelerate.

These observations of supernovae, which have been complemented by observations of the cosmic microwave background~\cite{Ade:2015xua} and baryon acoustic oscillations~\cite{Anderson:2013zyy,Alam:2016hwk}, have stipulated the development of a plethora of models. Possible explanations for the accelerating expansion include introducing a new type of fluid with barotropic index \(w < -1/3\) known as dark energy~\cite{Peebles:2002gy,Copeland:2006wr}, additional fields besides the metric~\cite{Faraoni:2004pi,Joyce:2014kja}, higher dimensional models~\cite{Dvali:2000hr,Lue:2005ya} or modifying the action of gravity~\cite{Sotiriou:2008rp,Capozziello:2011et}, possibly introducing a different description for the metric geometry of spacetime~\cite{Linder:2010py,Cai:2015emx}. However, despite this large theoretical effort the nature of dark energy and the cause of the accelerating expansion have so far remained undisclosed.

The aforementioned models have in common that the propagation of light and the tick rates of clocks, which are crucial ingredients to the calculation of the magnitude-redshift relation, are determined from the FLRW metric geometry of spacetime: light follows null geodesics of the metric and clocks measure the metric arc length along their world lines. However, the notions of null geodesics and arc length are defined not only in metric geometry, but also in more general geometries. The most general geometry to define the notion of arc length of a curve, which also defines the notion of geodesics, is Finsler geometry.

Finsler geometry is a straightforward mathematical generalization of Riemannian~\cite{BCS} and, after some refinements of its formulation, also of Lorentzian metric spacetime geometry~\cite{Asanov,Aazami:2014ata,Barletta2012,Beem,Caponio:2015hca,Hohmann:2013fca,javaloyes2014,Lammerzahl:2012kw,Minguzzi:2014aua,Perlick:2005hz,Pfeifer:2011tk}. It has been realized that Finsler geometry describes geometries of spacetimes which allow for deviations from Lorentz invariance in their local symmetries \cite{Gibbons:2007iu, Schreck:2015seb} and, moreover, the motion of test particles which obey modified dispersion relations. These may emerge as effective descriptions of the interaction of a quantized theory of gravity with test particles \cite{Amelino-Camelia:2014rga, Letizia:2016lew}, or in general from non-metric field theories \cite{Raetzel:2010je} like area metric, or general linear, electrodynamics \cite{Punzi:2006hy,Punzi:2006nx,Punzi:2007di} and effective field theories describing waves in media \cite{Cerveny,Klimes,Antonelli,Gibbons:2011ib,Perlick:2005hz}.

Despite this wide range of applications in physics a thorough and complete analysis of the impact of a Finslerian modification of the geometry of spacetime on astrophysical and gravitational observables is still missing. Several steps of such a phenomenological analysis have been done, however this program is far from being complete. In \cite{Lammerzahl:2012kw} a specific first order Finsler perturbation of Schwarzschild geometry and its effects on test particle motion has been analyzed, while in \cite{Pfeifer:2011xi} we showed how another class of spherically symmetric Finsler modification of Minkowski spacetime can address the fly-by anomaly in the solar system. Again another class of Finsler spacetime geometries has been studied towards its influence on gravitational waves \cite{Fuster:2015tua}. The effect of Finsler geometry on an observer's measurement of length has been studied in \cite{Pfeifer:2014yua} and consequences on the weak equivalence principle in \cite{Minguzzi:2016gct}. In addition to these studies on the influence of a Finsler modification of the geometry of spacetime on gravitational observables, the influence of a specific Finsler modification on the hydrogen atom has been investigated \cite{Itin:2014uia}.

Due to the fact that Finsler spacetime geometry is based on a homogeneous function on the tangent bundle of spacetime, instead of on a tensor field such as a metric, it is difficult to analyze observable effects for general Finsler modifications of the geometry of spacetime. Therefore the observable consequences analyzed in the articles mentioned usually choose a specific Finsler spacetime model to perform their studies. In our analysis in this article we will be keeping the maximal degree of generality whenever possible. However, when we want to derive explicit observable consequences, we need to choose specific Finsler spacetime models to make predictions. The long term goal is to find a systematic scheme to analyze observables of a Finslerian spacetime modification and their deviation from metric spacetime geometry, in a framework similar to the parametrized post-Newtonian formalism.

In this article we consider Finsler spacetimes with cosmological symmetry which are based on the construction we developed in the articles~\cite{Pfeifer:2011tk,Pfeifer:2011xi,Pfeifer:2013gha,Hohmann:2013fca,Hohmann:2014gpa}. The central goal of our work is to derive the magnitude-redshift relation, and thus also the deceleration parameter \(q\), under the assumption of a cosmological Finsler background geometry. For this purpose we study the geodesic motion of both massive test bodies and in particular light in this background geometry. These studies yield us the geodesic equation, which can most conveniently be expressed by a vector field on the tangent bundle called the geodesic spray, and a full set of constants of motion. As a second ingredient we use the tick rates of co-moving clocks on cosmological Finsler spacetimes, in order to compare the frequencies of emitted and observed photons. Having derived a general equation for the magnitude-redshift relation on general homogeneous and isotropic Finsler spacetimes we reach the point where we need to consider specific models to obtain observable predictions. We choose several classes of Finsler spacetime geometries, which can be used as generalizations of Lorentzian metric spacetime geometry in physics, and display expressions for their deceleration parameters.

The outline of this article is as follows. In section~\ref{sec:finslercosmo} we briefly review the notion of Finsler spacetimes with cosmological symmetry. We then discuss geodesic motion on cosmological Finsler spacetimes in section~\ref{sec:geodmotion}. We derive a general formula for the magnitude-redshift relation in section~\ref{sec:magred}. In section~\ref{sec:examples} we apply our findings to a number of examples. We end with a discussion in section~\ref{sec:discussion}. Lengthy formulas are displayed in a number of appendices: these are in particular the complete lifts of the cosmological symmetry generators in appendix~\ref{app:CL}, the geodesic spray in appendix~\ref{app:geodspray} and the radial geodesics in appendix~\ref{app:radialgeod}.

%%%%%%%%%%%%%%%%%%%%%%%%%%%%%%%%%%%%%%%%%%%%%%%%%%%%%%%%%%%%%%%%%%%%%%%%%%%%%%%%%%
\section{Cosmological Finsler spacetimes}\label{sec:finslercosmo}
In order to derive the cosmological redshift on homogeneous and isotropic Finsler spacetimes we start by introducing the mathematical notations we need during the remainder of this article. For this purpose we briefly review the definition of Finsler spacetimes in section~\ref{ssec:finslerst}. We then display the generators of cosmological symmetry in section~\ref{ssec:cosmosym}. For convenience, we finally introduce adapted coordinates on the tangent bundle in section~\ref{ssec:cosmocoord}, which will simplify our calculations. Further mathematical details and the derivation of the most general cosmological Finsler spacetime can be found in the articles \cite{Pfeifer:2011tk, Pfeifer:2011xi, Hohmann:2015pva, Hohmann:2015ywa} and in the thesis \cite{Pfeifer:2013gha}.

%------------------------------------------------------------------------------------------%
\subsection{Finsler spacetimes}\label{ssec:finslerst}
Finsler spacetimes are straightforward generalizations of Lorentzian metric spacetimes. Instead of a metric which defines the geometry of a spacetime $M$ one derives the geometry from a general length measure for curves on $M$. This concept was developed by Finsler in 1918 \cite{finsler} and was further developed by many authors since then. For the application in physics it is important to deal with indefinite length measures to distinguish between timelike, lightlike and spacelike curves. To incorporate these notions into Finsler geometry one of us developed the Finsler spacetime framework \cite{Pfeifer:2013gha} which extends and includes a previous approach to indefinite Finsler geometry by Beem \cite{Beem}.

Finsler spacetime geometry is formulated on the tangent bundle $TM$ of the spacetime $M$. We use the following notations. An element of the tangent bundle $Z\in TM$ is a vector in some tangent space $T_xM$ to the spacetime manifold. In local coordinates $x$ in $M$ we can write $Z= {y^a\partial_a}_{|x}$ where $y^a$ are the components of the vector $Z$ with respect to the coordinate basis of $T_xM$. This means we can label the point $Z$ on the tangent bundle with the coordinates $(x,y)$, which are called manifold induced coordinates of the tangent bundle. The corresponding coordinate basis of the tangent spaces of the tangent bundle will be labeled by $\frac{\partial}{\partial x^a}=\partial_a$ and $\frac{\partial}{\partial y^a} = \bar{\partial}_a$ and its co-basis is denoted by $dx^a$ and $dy^a$.

The precise definition of a Finsler spacetime we use is the one developed in \cite{Pfeifer:2011tk}.

\vspace{6pt}\noindent\textbf{Definition 1.}
\textit{A Finsler spacetime $(M,L)$ is a four-dimensional, connected, Hausdorff, paracompact, smooth manifold~$M$ equipped with a continuous function $L:TM\rightarrow\mathbb{R}$ on the tangent bundle which has the following properties:
\begin{enumerate}[(i)]
	\item $L$ is smooth on the tangent bundle without the zero section $TM\setminus\{0\}$;\vspace{-6pt}
	\item $L$ is positively homogeneous of real degree $h\ge 2$ with respect to the fiber coordinates of $TM$,
	\begin{equation}\label{eqn:hom}
	L(x,\lambda y) = \lambda^h L(x,y) \quad \forall \lambda>0\,;
	\end{equation}
	\item \vspace{-6pt}$L$ is reversible in the sense
	\begin{equation}\label{eqn:rev}
	|L(x,-y)|=|L(x,y)|\,;
	\end{equation}
	\item \vspace{-6pt}the Hessian $g^L_{ab}$ of $L$ with respect to the fiber coordinates is non-degenerate on $TM\setminus A$ where~$A$ has measure zero and does not contain the null set $\{(x,y)\in TM\,|\,L(x,y)=0\}$,
	\begin{equation}
	g^L_{ab}(x,y) = \frac{1}{2}\bar\partial_a\bar\partial_b L\,;
	\end{equation}
	\item \vspace{-6pt}the unit timelike condition holds, i.e., for all $x\in M$ the set
	\begin{equation}
	\Omega_x=\Big\{y\in T_xM\,\Big|\, |L(x,y)|=1\,,\;g^L_{ab}(x,y)\textrm{ has signature }(\epsilon,-\epsilon,-\epsilon,-\epsilon)\,,\, \epsilon=\frac{|L(x,y)|}{L(x,y)}\Big\}
	\end{equation}
	contains a non-empty closed connected component $S_x\subset \Omega_x\subset T_xM$.
\end{enumerate}
The Finsler function associated to $L$ is $F(x,y) = |L(x,y)|^{1/h}$ and the Finsler metric $g^F_{ab}=\frac{1}{2}\bar\partial_a \bar\partial_b F^2$.}

Basically this very general definition of Finsler spacetimes ensures that the Finsler spacetime geometry allows for a precise notion of timelike, lightlike and spacelike directions as well as for a well defined geometry on the null-structure of the $L$, along which light rays propagate, and along all timelike directions.

%------------------------------------------------------------------------------------------%
\subsection{Homogeneity and isotropy}\label{ssec:cosmosym}
A symmetry of a Finsler spacetime is defined by vector fields $X=\xi^a\partial_a$ on spacetime whose complete lifts $X^C = \xi^a\partial_a + y^m\partial_m\xi^a\bar\partial_a$ annihilate the fundamental geometry function
\begin{align}
X^C(L) = 0\,.
\end{align}
For cosmological symmetry we start in spherical coordinates $(t,r,\theta,\phi)$ on $M$ and the corresponding manifold induced coordinates $(t,r,\theta,\phi, y^t, y^r, y^\theta, y^\phi)$ on $TM$. The generators of the cosmological symmetry, i.e., homogeneity and isotropy, are the generators of rotations
\begin{subequations}\label{eqn:rotgen}
\begin{align}
\rho_1 &= \sin\phi\partial_{\theta} + \frac{\cos\phi}{\tan\theta}\partial_{\phi}\,,\\
\rho_2 &= -\cos\phi\partial_{\theta} + \frac{\sin\phi}{\tan\theta}\partial_{\phi}\,,\\
\rho_3 &= \partial_{\phi}
\end{align}
\end{subequations}
and translations
\begin{subequations}\label{eqn:transgen}
\begin{align}
\tau_1 &= \sqrt{1 - kr^2}\left(\sin\theta\cos\phi\partial_r + \frac{\cos\theta\cos\phi}{r}\partial_{\theta} - \frac{\sin\phi}{r\sin\theta}\partial_{\phi}\right)\,,\\
\tau_2 &= \sqrt{1 - kr^2}\left(\sin\theta\sin\phi\partial_r + \frac{\cos\theta\sin\phi}{r}\partial_{\theta} + \frac{\cos\phi}{r\sin\theta}\partial_{\phi}\right)\,,\\
\tau_3 &= \sqrt{1 - kr^2}\left(\frac{\sin\theta}{r}\partial_{\theta} - \cos\theta\partial_r\right)\,.
\end{align}
\end{subequations}
Their complete lifts to the tangent bundle are listed in the appendix \ref{app:CL}. Demanding that the lifted vector fields annihilate the fundamental geometry function, as described above, yields that the fundamental geometry function must be of the form
\begin{equation}\label{eqn:lcosmind}
L(x,y) = L(t,y^t,w(r,\theta,\phi,y^r,y^\theta,y^\phi)) \text{ with } w^2 = \frac{(y^r)^2}{1-kr^2}+ r^2 (y^\theta)^2 + r^2 \sin^2\theta (y^\phi)^2\,.
\end{equation}
However, working in these coordinates turns out to be rather cumbersome. In the following we therefore make use of a more convenient set of coordinates on the tangent bundle.

%------------------------------------------------------------------------------------------%
\subsection{Adapted Coordinates}\label{ssec:cosmocoord}
For the analysis of timelike geodesics, it turns out to be useful to introduce coordinates \(y, u, v, w\) on each tangent space such that
\begin{equation}\label{eqn:coorddef}
y^t = y\,, \quad y^r = w\cos u\sqrt{1 - kr^2}\,, \quad y^{\theta} = \frac{w}{r}\sin u\cos v\,, \quad y^{\phi} = \frac{w}{r\sin\theta}\sin u\sin v\,.
\end{equation}
In these coordinates the complete lifts of the generators of cosmological symmetry take the form
\begin{subequations}\label{eqn:rotgenc}
\begin{align}
\rho^C_1 &= \sin\phi\partial_{\theta} + \frac{\cos\phi}{\tan\theta}\partial_{\phi} - \frac{\cos\phi}{\sin\theta}\partial_v\,,\\
\rho^C_2 &= -\cos\phi\partial_{\theta} + \frac{\sin\phi}{\tan\theta}\partial_{\phi} - \frac{\sin\phi}{\sin\theta}\partial_v\,,\\
\rho^C_3 &= \partial_{\phi}
\end{align}
\end{subequations}
and
\begin{subequations}\label{eqn:transgenc}
\begin{align}
\tau^C_1 &= \sqrt{1 - kr^2}\left(\sin\theta\cos\phi\partial_r + \frac{\cos\theta\cos\phi}{r}\partial_{\theta} - \frac{\sin\phi}{r\sin\theta}\partial_{\phi}\right) - \frac{\cos v\cos\theta\cos\phi - \sin v\sin\phi}{r}\partial_u\nonumber\\
&\phantom{=}+ \frac{\sin v\sin\theta\cos\phi + \cos v\tan\theta\sin\phi + \tan u\sin\phi\sqrt{1 - kr^2}}{r\tan u\tan\theta}\partial_v\,,\\
\tau^C_2 &= \sqrt{1 - kr^2}\left(\sin\theta\sin\phi\partial_r + \frac{\cos\theta\sin\phi}{r}\partial_{\theta} + \frac{\cos\phi}{r\sin\theta}\partial_{\phi}\right) - \frac{\cos v\cos\theta\sin\phi + \sin v\cos\phi}{r}\partial_u\nonumber\\
&\phantom{=}+ \frac{\sin v\sin\theta\sin\phi - \cos v\tan\theta\cos\phi - \tan u\cos\phi\sqrt{1 - kr^2}}{r\tan u\tan\theta}\partial_v\,,\\
\tau^C_3 &= \sqrt{1 - kr^2}\left(\frac{\sin\theta}{r}\partial_{\theta} - \cos\theta\partial_r\right) + \frac{\sin\theta}{r}\left(\frac{\sin v}{\tan u}\partial_v - \cos v\partial_u\right)\,.
\end{align}
\end{subequations}
which is significantly simpler than the corresponding expressions shown in appendix~\ref{app:CL}. One now easily verifies that the most general cosmologically symmetric Finsler geometry function reads
\begin{equation}
L = L(t, y, w) = y^h\tilde{L}(t, w/y)\,.
\end{equation}
The second equality holds wherever $y\neq 0$ and follows from the fact that \(L\) is homogeneous of degree~\(h\). The dependence only on $(t,y,w)$ is consistent with the expression~\eqref{eqn:lcosmind} in induced coordinates.

In the following we will study future timelike curves. The tangent vectors of these curves constitute the interior of the forward light cone and hence satisfy \(y > 0\) and \(\tilde{L} \neq 0\) \footnote{Observe that we do not fix the sign of $\tilde L$ here since there exist interesting examples with either sign of $L$ in the interior of the forward light cone; see the FLRW and the Randers example in section \ref{sec:examples}.}. In the interior of the forward light cone we now introduce another set of convenient coordinates:
\begin{equation}\label{eq:adapted}
T = t\,, \quad R = r\,, \quad \Theta = \theta\,, \quad \Phi = \phi\,, \quad Y = y^h\tilde{L}\left(t, \frac{w}{y}\right)\,, \quad U = u\,, \quad V = v\,, \quad W = \frac{w}{y}\,,
\end{equation}
or conversely,
\begin{equation}\label{eq:invadapted}
t = T\,, \quad r = R\,, \quad \theta = \Theta\,, \quad \phi = \Phi\,, \quad y = \left(\frac{Y}{\tilde{L}(T, W)}\right)^{\frac{1}{h}}\,, \quad u = U\,, \quad v = V\,, \quad w = W\left(\frac{Y}{\tilde{L}(T, W)}\right)^{\frac{1}{h}}\,.
\end{equation}
Note that these coordinates would become singular at \(y = 0\) and \(\tilde{L} = 0\) and can therefore not be used on the null structure and for the analysis of null-geodesics, hence their restricted domain. However, from the fact that \(y > 0\) on this domain follows that $Y$ and $\tilde L$ always have the same sign, so that both the transformation~\eqref{eq:adapted} and its inverse~\eqref{eq:invadapted} are well-defined and differentiable, and hence constitute a viable coordinate transformation. We further introduce the notation
\begin{equation}
\tilde{L}_t = \partial_T\tilde{L}\,, \quad \tilde{L}_w = \partial_W\tilde{L}
\end{equation}
for the derivatives of \(\tilde{L}\) with respect to its first and second argument. We also have
\begin{equation}
\partial_t\tilde{L} = \tilde{L}_t\,, \quad \partial_w\tilde{L} = \frac{\tilde{L}_w}{y}\,, \quad \partial_y\tilde{L} = -\frac{w\tilde{L}_w}{y^2}\,,
\end{equation}
which we will use frequently in the following section, where we discuss geodesic motion.

%%%%%%%%%%%%%%%%%%%%%%%%%%%%%%%%%%%%%%%%%%%%%%%%%%%%%%%%%%%%%%%%%%%%%%%%%%%%%%%%%%
\section{Geodesic Motion}\label{sec:geodmotion}
We now come to the discussion of geodesic motion on cosmologically symmetric Finsler spacetimes as described in the preceding section. Recall that geodesics are conventionally defined as curves which are extremal with respect to a length functional. We briefly review the Finsler length functional and its relation to the geodesic equation in section~\ref{ssec:geodeqn}. The geodesic equation can be expressed in terms of a vector field \(\mathbf{S}\) on the tangent bundle, which we derive in the cosmologically symmetric case in section~\ref{ssec:geodspray}. Functions on the tangent bundle which are constant along the integral curves of the geodesic spray are constants of motion, and we display them in section~\ref{ssec:consmot}. We show their completeness in section~\ref{ssec:geodrec} by reconstructing the geodesic equation from the constants of motion. Finally, we give explicit expressions for radial geodesics in our cosmological coordinates in section~\ref{ssec:radialgeo}.

%------------------------------------------------------------------------------------------%
\subsection{The geodesic equation}\label{ssec:geodeqn}
The length measure for a curve \(\gamma\) on a Finsler spacetime, which is also the action for the motion of point particles, is given by
\begin{align}\label{eq:length}
S[\gamma] = \int d\tau F(\gamma,\dot{\gamma})\,.
\end{align}
Free particles follow the geodesics of this length functional. Due to the homogeneity of \(F\) of degree~1 it is invariant under a change of parametrization, so that its Euler-Lagrange equations cannot be brought into the form $\ddot x + G = 0$, since the bilinear form defined by the second derivative of \(F\), not $F^2$ or $L$, is degenerate. One has to fix the parametrization of the curves to $F(\gamma, \dot{\gamma})=\text{const}$. Then the Euler-Lagrange equations become
\begin{align}
	\ddot x^a + G^a(x,\dot x) = 0\,.
\end{align}
The functions $G^a(x,\dot x)$ define a vector field on $TM$, the so-called geodesic spray $\mathbf{S} = y^a\partial_a - G^a\bar{\partial}_a$, whose integral curves are the geodesics. Physically free particles propagate through spacetime along such geodesics which have tangents that are either null $F(\gamma, \dot{\gamma})=0$ or belong to the cone of future timelike vectors, which exists by the definition of Finsler spacetimes. To calculate the geodesic spray in manifold induced coordinates is a quite lengthy task and is displayed in appendix \ref{app:geodspray}. In the following section we display the geodesic spray in cosmological coordinates, in which it takes a more compact form.

%------------------------------------------------------------------------------------------%
\subsection{The geodesic spray}\label{ssec:geodspray}
In arbitrary, non-induced coordinates on the tangent bundle we can calculate the geodesic spray \(\mathbf{S}\) as follows. We start with the differential \(dL\) of the geometry function \(L\), which in the cosmological case reads
\begin{equation}
dL = y^h\tilde{L}_t dt + y^{h - 2}(hy\tilde{L} - w\tilde{L}_w)dy + y^{h - 1}\tilde{L}_w dw\,.
\end{equation}
Together with the cotangent structure \(J^*\), which can be written in manifold induced coordinates as \(J^* = dx^a \otimes \bar{\partial}_a\), this yields the Cartan one-form
\begin{equation}\label{eqn:cartanone}
\theta^L = J^*dL = y^{h - 2}(hy\tilde{L} - w\tilde{L}_w)dt + y^{h - 1}\tilde{L}_w\left[\frac{\cos u}{\sqrt{1 - kr^2}}dr + r\sin u\left(\cos v\,d\theta + \sin v\sin\theta\,d\phi\right)\right]\,.
\end{equation}
Its exterior derivative \(\omega^L = d\theta^L\) is a symplectic form on \(TM\setminus A\), called the Cartan two-form. Hence, there exists a unique vector field \(\mathbf{S}\) such that
\begin{equation}
\iota_{\mathbf{S}}\omega^L = -(h - 1)dL\,.
\end{equation}
This vector field is the geodesic spray. In cosmological coordinates on the tangent bundle it reads
\begin{multline}\label{eq:SfromF}
	\mathbf{S} = y\partial_t + w\cos u\sqrt{1 - kr^2}\partial_r + \frac{w\sin u\cos v}{r}\partial_{\theta} + \frac{w\sin u\sin v}{r\sin\theta}\partial_{\phi} - \frac{w\sin u\sqrt{1 - kr^2}}{r}\partial_u\\
	- \frac{w\sin u\sin v}{r\tan\theta}\partial_v - y^2\frac{\tilde{L}_{ww}\tilde{L}_t - \tilde{L}_w\tilde{L}_{tw}}{h\tilde{L}\tilde{L}_{ww} - (h - 1)\tilde{L}_w^2}\partial_y - y\frac{w\tilde{L}_{t}\tilde{L}_{ww} + hy\tilde{L}\tilde{L}_{tw} - w\tilde{L}_w\tilde{L}_{tw} - (h - 1)y\tilde{L}_t\tilde{L}_w}{h\tilde{L}\tilde{L}_{ww} - (h - 1)\tilde{L}_w^2}\partial_w\,.
\end{multline}
In observer coordinates it takes the simpler form
\begin{multline}
\mathbf{S} = \left(\frac{Y}{\tilde{L}}\right)^{\frac{1}{h}}\Bigg(\partial_T + W\cos U\sqrt{1 - kR^2}\partial_R + \frac{W\sin U\cos V}{R}\partial_{\Theta} + \frac{W\sin U\sin V}{R\sin\Theta}\partial_{\Phi}\\
	- \frac{W\sin U\sqrt{1 - kR^2}}{R}\partial_U - \frac{W\sin U\sin V}{R\tan\Theta}\partial_V - \frac{h\tilde{L}\tilde{L}_{tw} - (h - 1)\tilde{L}_t\tilde{L}_w}{h\tilde{L}\tilde{L}_{ww} - (h - 1)\tilde{L}_w^2}\partial_W\Bigg)\,.
\end{multline}
We will make use of these expressions later when we apply the geodesic equation to the motion of test bodies and light.

%------------------------------------------------------------------------------------------%
\subsection{Constants of motion}\label{ssec:consmot}
If \(X = \xi^a\partial_a\) is a vector field on \(M\) generating a symmetry of the Finsler spacetime, so that the complete lift satisfies \(X^CL = 0\), then there exists a function \(C_{X} = \iota_{X^C}\theta^L\) on \(TM\) which is constant along geodesics, \(\mathbf{S}C_{X} = 0\). In manifold induced coordinates this formula translates into $C_X = \xi^a\bar{\partial}_a L$.

In order to calculate the constants of motion on a cosmologically symmetric Finsler spacetime we can make use of the expression~\eqref{eqn:cartanone} for the Cartan one-form and~\eqref{eqn:rotgenc} and~\eqref{eqn:transgenc} for the complete lifts of the symmetry generating vector fields. For the generators~\eqref{eqn:rotgen} of rotations we then obtain the angular momentum
\begin{subequations}
	\begin{align}
		\Lambda_1 &= \tilde{L}_wry^{h - 1}\sin u(\sin v\cos\theta\cos\phi + \cos v\sin\phi)\,,\\
		\Lambda_2 &= \tilde{L}_wry^{h - 1}\sin u(\sin v\cos\theta\sin\phi - \cos v\cos\phi)\,,\\
		\Lambda_3 &= \tilde{L}_wry^{h - 1}\sin u\sin v\sin\theta\,,
	\end{align}
\end{subequations}
while for the generators~\eqref{eqn:transgen} of translations we have the linear momentum
\begin{subequations}
	\begin{align}
		\Pi_1 &= \tilde{L}_wy^{h - 1}\left[\sin u\left(\cos v\cos\theta\cos\phi - \sin v\sin\phi\right)\sqrt{1 - kr^2} + \cos u\sin\theta\cos\phi\right]\,,\\
		\Pi_2 &= \tilde{L}_wy^{h - 1}\left[\sin u\left(\cos v\cos\theta\sin\phi + \sin v\cos\phi\right)\sqrt{1 - kr^2} + \cos u\sin\theta\sin\phi\right]\,,\\
		\Pi_3 &= \tilde{L}_wy^{h - 1}\left(\sin u\cos v\sin\theta\sqrt{1 - kr^2} - \cos u\cos\theta\right)\,.
	\end{align}
\end{subequations}
We also use the shorthand notations \(\vec{\Lambda}\) and \(\vec{\Pi}\). Note that these are not independent, but satisfy \(\vec{\Lambda} \cdot \vec{\Pi} = 0\). Also, \(C_0 = L = y^h\tilde{L}\) always is a constant of motion. Since the expressions above are rather lengthy, it is useful to express them in terms of simpler expressions, which can be constructed from the original ones. From the squared vectors
\begin{equation}
	\vec{\Lambda}^2 = \tilde{L}_w^2r^2y^{2h - 2}\sin^2u\,, \quad \vec{\Pi}^2 = \tilde{L}_w^2y^{2h - 2}\left(1 - kr^2\sin^2u\right)
\end{equation}
we can construct
\begin{equation}
	C_1^2 = \vec{\Pi}^2 + k\vec{\Lambda}^2 = y^{2h - 2}\tilde{L}_w^2\,, \quad C_2^2 = \frac{\vec{\Lambda}^2}{\vec{\Pi}^2 + k\vec{\Lambda}^2} = r^2\sin^2u\,.
\end{equation}
By making use of the relations
\begin{equation}
	\frac{\Lambda_1}{\Lambda_2} = \frac{\tan v\cos\theta + \tan\phi}{\tan v\cos\theta\tan\phi - 1} = -\tan[\phi + \arctan(\tan v\cos\theta)]\,, \quad \frac{\Lambda_3^2}{\vec{\Lambda}^2} = \sin^2v\sin^2\theta
\end{equation}
one can further read off the constants of motion
\begin{equation}
	C_3 = \phi + \arctan(\tan v\cos\theta)\,, \quad C_4 = \sin v\sin\theta\,.
\end{equation}
Finally, we can define
\begin{subequations}
	\begin{align}
		C_5 &= \frac{\Pi_3}{C_1} = \sin u\cos v\sin\theta\sqrt{1 - kr^2} - \cos u\cos\theta\,,\\
		C_6 &= \frac{\Lambda_1\Pi_2 - \Pi_1\Lambda_2}{C_1^2C_2} = \sin u\cos\theta\sqrt{1 - kr^2} + \cos u\cos v\sin\theta\,.
	\end{align}
\end{subequations}
Making use of these formulas, we can now fully express \(L, \vec{\Lambda}, \vec{\Pi}\) in terms of the constants \(C_0, \ldots, C_6\). First, note that
\begin{equation}
	L = C_0\,, \quad \vec{\Lambda}^2 = C_1^2C_2^2\,, \quad \vec{\Pi}^2 = C_1^2(1 - kC_2^2)\,.
\end{equation}
One can then easily read off the third components
\begin{equation}
	\Lambda_3 = C_1C_2C_4\,, \quad \Pi_3 = C_1C_5\,.
\end{equation}
Finally, the remaining components are given by
\begin{align}
	\Lambda_1 &= C_1C_2\sin C_3\sqrt{1 - C_4^2}\,, & \Pi_1 &= \frac{C_1(C_6\cos C_3 - C_4C_5\sin C_3)}{\sqrt{1 - C_4^2}}\,,\\
	\Lambda_2 &= -C_1C_2\cos C_3\sqrt{1 - C_4^2}\,, & \Pi_2 &= \frac{C_1(C_6\sin C_3 + C_4C_5\cos C_3)}{\sqrt{1 - C_4^2}}\,.
\end{align}
Of course, also the constants \(C_0, \ldots, C_6\) are not independent, since they are related by
\begin{equation}
	1 = \frac{\vec{\Pi}^2}{C_1^2} + kC_2^2 = \frac{C_5^2 + C_6^2}{1 - C_4^2} + kC_2^2\,.
\end{equation}
The constants of motion form a complete set in the sense that they fully determine the geodesic equation, and hence the geodesic spray, as will see in the following.

%------------------------------------------------------------------------------------------%
\subsection{Reconstruction of the geodesic equation from constants of motion}\label{ssec:geodrec}
We now show that the coefficients \(G^a\) in the geodesic spray \(\mathbf{S} = y^a\partial_a - G^a\bar{\partial}_a\) can also be obtained from the constants of motion shown above. By making use of the definition~\eqref{eqn:coorddef} of the adapted coordinates, we can express the geodesic spray by an ansatz of the form
\begin{equation}
\mathbf{S} = y\partial_t + w\cos u\sqrt{1 - kr^2}\partial_r + \frac{w\sin u\cos v}{r}\partial_{\theta} + \frac{w\sin u\sin v}{r\sin\theta}\partial_{\phi} - G^y\partial_y - G^u\partial_u - G^v\partial_v - G^w\partial_w\,,
\end{equation}
where \(G^y, G^u, G^v, G^w\) are to be determined from the constants of motion. From this ansatz we obtain the linear system
\begin{subequations}
\begin{align}
0 = \mathbf{S}C_0 &= y^{h - 2}\left[y^3\tilde{L}_t - hy\tilde{L}G^y + (wG^y - yG^w)\tilde{L}_w\right]\,,\\
0 = \mathbf{S}C_1 &= y^{h - 3}\left[y^3\tilde{L}_{tw} - (h - 1)y\tilde{L}_wG^y + (wG^y - yG^w)\tilde{L}_{ww}\right]\,,\\
0 = \mathbf{S}C_2 &= \left(w\sin u\sqrt{1 - kr^2} - rG^u\right)\cos u\,,\\
0 = \mathbf{S}C_4 &= \left(\frac{w\sin u\sin v\cos\theta}{r} - \sin\theta\,G^v\right)\cos v\,,
\end{align}
\end{subequations}
which is easily solved by
\begin{gather}
G^u = \frac{w\sin u\sqrt{1 - kr^2}}{r}\,, \quad
G^v = \frac{w\sin u\sin v}{r\tan\theta}\,,\nonumber\\
G^y = y^2\frac{\tilde{L}_{ww}\tilde{L}_t - \tilde{L}_w\tilde{L}_{tw}}{h\tilde{L}\tilde{L}_{ww} - (h - 1)\tilde{L}_w^2}\,, \quad
G^w = y\frac{w\tilde{L}_{t}\tilde{L}_{ww} + hy\tilde{L}\tilde{L}_{tw} - w\tilde{L}_w\tilde{L}_{tw} - (h - 1)y\tilde{L}_t\tilde{L}_w}{h\tilde{L}\tilde{L}_{ww} - (h - 1)\tilde{L}_w^2}\,.
\end{gather}
One can see immediately that this agrees with the result~\eqref{eq:SfromF}.

%------------------------------------------------------------------------------------------%
\subsection{Radial Geodesics}\label{ssec:radialgeo}
We consider in particular radial geodesics, for which the angles \(\theta \equiv \pi/2\) and \(\phi \equiv 0\) are constant, so that the geodesic is specified by functions \(t(\lambda)\) and \(r(\lambda)\), if we allow for arbitrary parametrizations. One of these functions will be fixed by the choice of the parametrization. The canonical lift of such a geodesic to the tangent bundle then has
\begin{equation}
	y^t = \dot{t}\,, \quad y^{\theta} = \dot{\theta}= 0\,, \quad y^{\phi} = \dot{\phi} = 0\,, \quad y^r = \dot{r}\,,
\end{equation}
where dots denote derivatives with respect to \(\lambda\). One could now make use of the geodesic equation in manifold induced coordinates given in appendix~\ref{app:geodspray}; this procedure is detailed in appendix~\ref{app:radialgeod}. However, it turns out to be easier to rewrite the left hand sides in the equations above, which are given by induced coordinates, in terms of the adapted coordinates. We then obtain
\begin{equation}
	y = \dot{t}\,, \quad u = 0\,, \quad v = 0\,, \quad w\sqrt{1 - kr^2} = \dot{r}\,.
\end{equation}
In this case the constants of motion derived in the previous section are given by
\begin{equation}\label{eqn:radconsmot}
C_0 = y^h\tilde{L}\,, \quad C_1 = y^{h - 1}\tilde{L}_w\,, \quad C_6 = 1\,, \quad C_2 = C_3 = C_4 = C_5 = 0\,.
\end{equation}
Moreover the geodesics must be integral curves of the geodesic spray in order to satisfy the geodesic equation. Hence, they must further satisfy the relations
\begin{equation}
	\dot{y} = -y^2\frac{\tilde{L}_{ww}\tilde{L}_t - \tilde{L}_w\tilde{L}_{tw}}{h\tilde{L}\tilde{L}_{ww} - (h - 1)\tilde{L}_w^2}\,, \quad \dot{w} = -y\frac{w\tilde{L}_{t}\tilde{L}_{ww} + hy\tilde{L}\tilde{L}_{tw} - w\tilde{L}_w\tilde{L}_{tw} - (h - 1)y\tilde{L}_t\tilde{L}_w}{h\tilde{L}\tilde{L}_{ww} - (h - 1)\tilde{L}_w^2}\,.
\end{equation}
These two equations can also be derived using the constants of motion \(C_0\) and \(C_1\) shown in equation~\eqref{eqn:radconsmot}. From their derivative with respect to the curve parameter follows
\begin{subequations}
\begin{align}
0 &= \frac{dC_0}{d\lambda} = y^h\left[\tilde{L}_t\dot{t} + \frac{hy\tilde{L} - w\tilde{L}_w}{y^2}\dot{y} + \frac{\tilde{L}_w}{y}\dot{w}\right]\,,\\
0 &= \frac{dC_1}{d\lambda} = y^h\left[\frac{\tilde{L}_{tw}}{y}\dot{t} + \frac{(h - 1)y\tilde{L}_w - w\tilde{L}_{ww}}{y^3}\dot{y} + \frac{\tilde{L}_{ww}}{y^2}\dot{w}\right]\,.
\end{align}
\end{subequations}
Inserting \(\dot{t} = y\) and solving the resulting linear system for \(\dot{y}\) and \(\dot{w}\) then yields the geodesic equation as shown above.

In the case of a timelike geodesic we can rewrite the geodesic equation also in observer coordinates, so that it takes the simpler form
\begin{equation}
	\dot{T} = \left(\frac{Y}{\tilde{L}}\right)^{\frac{1}{h}}\,, \quad \dot{R} = \left(\frac{Y}{\tilde{L}}\right)^{\frac{1}{h}}W\sqrt{1 - kR^2}\,, \quad \dot{Y} = 0\,, \quad \dot{W} = -\left(\frac{Y}{\tilde{L}}\right)^{\frac{1}{h}}\frac{h\tilde{L}\tilde{L}_{tw} - (h - 1)\tilde{L}_t\tilde{L}_w}{h\tilde{L}\tilde{L}_{ww} - (h - 1)\tilde{L}_w^2}\,.
\end{equation}
It is obvious that this rewriting procedure into these coordinates is not possible for null-geodesics, due to the appearance of a factor $\tilde{L}$ in the denominator. One can now use the fact that \(\dot{Y} = 0\) and consider the special case of a geodesic in arc length parametrization \(Y = 1\). In this case the geodesic equation reduces to
\begin{equation}
	\dot{T} = \tilde{L}^{-\frac{1}{h}}\,, \quad \dot{R} = \tilde{L}^{-\frac{1}{h}}W\sqrt{1 - kR^2}\,, \quad \dot{W} = -\tilde{L}^{-\frac{1}{h}}\frac{h\tilde{L}\tilde{L}_{tw} - (h - 1)\tilde{L}_t\tilde{L}_w}{h\tilde{L}\tilde{L}_{ww} - (h - 1)\tilde{L}_w^2}\,.
\end{equation}
The equations derived here will be the crucial ingredient for our derivation of the magnitude-redshift relation in the following section.

%%%%%%%%%%%%%%%%%%%%%%%%%%%%%%%%%%%%%%%%%%%%%%%%%%%%%%%%%%%%%%%%%%%%%%%%%%%%%%%%%%
\section{Magnitude-redshift relation}\label{sec:magred}
We can now use our results on the geodesic motion in a cosmologically symmetric Finsler spacetime detailed in the previous section in order to derive the magnitude-redshift relation. This will be done in three steps. First, we will calculate the redshift of a light source in section~\ref{ssec:redshift}. We will then calculate its observed magnitude in section~\ref{ssec:magnitude}. Relating these quantities will then yield us a series expansion of the magnitude-redshift relation in section~\ref{ssec:magred}. The leading order expansion coefficient, which is related to the deceleration parameter, allows for a comparison of the Finsler background geometry to observations. Since we do not fix any particular parametrization for the cosmological time coordinate, we will finally show the independence of our result from this choice in section~\ref{ssec:diffinv}.

%------------------------------------------------------------------------------------------%
\subsection{Redshift of a light source}\label{ssec:redshift}
We consider the emission of light at time \(t_e\) from a source located at cosmological coordinates \((r_e, \theta_e = \pi/2, \phi_e = 0)\). The light will be received by an observer at coordinates \((r_o, \theta_o = \pi/2, \phi_o = 0)\) at time \(t_o\). These two events must be connected by a lightlike radial geodesic, which we parametrize with curve parameter \(\lambda\). This allows us to write
\begin{equation}
\frac{dr}{dt} = \frac{\dot{r}}{\dot{t}} = \frac{w}{y}\sqrt{1 - kr^2}\,.
\end{equation}
Along the canonical lift of this geodesic we have that \(C_0 = y^h\tilde{L} \equiv 0\) is constant, since \(L\) is constant along geodesics. From \(y = \dot{t} > 0\) hence follows that \(\tilde{L} \equiv 0\) is also constant along the geodesic, so that we can determine \(W = w/y\) as a function of \(t\) by solving
\begin{equation}\label{eqn:geodredshift}
\tilde L(t, W):= L\left(t, 1, \frac{w}{y}\right) = 0
\end{equation}
for all \(t\). We can then use the solution, which we denote by \(\mathring{W}(t)\), to integrate
\begin{equation}
\int_{r_e}^{r_o}\frac{dr}{\sqrt{1 - kr^2}} = \int_{t_e}^{t_o}\mathring{W}(t)\,dt\,.
\end{equation}
Note that the integral on the left hand side only depends on the location of the source and the observer and is independent of the time when the signal was emitted and observed. If two subsequent periods of a wave are emitted at times \(t_{e,1}\) and \(t_{e,2}\) and observed at times \(t_{o,1}\) and \(t_{o,2}\), we thus have
\begin{equation}
0 = \int_{t_{e,2}}^{t_{o,2}}\mathring{W}(t)dt - \int_{t_{e,1}}^{t_{o,1}}\mathring{W}(t)dt = \int_{t_{o,1}}^{t_{o,2}}\mathring{W}(t)dt - \int_{t_{e,1}}^{t_{e,2}}\mathring{W}(t)dt \approx \mathring{W}(t_o)\Delta t_o - \mathring{W}(t_e)\Delta t_e\,,
\end{equation}
where we have first cut out the common integration domain from \(t_{e,2}\) to \(t_{o,1}\) and then used the physical assumption that \(\mathring{W}(t)\) does not change significantly within one period of radiation. This allows us to write the ratio of the coordinate time intervals as
\begin{equation}
\frac{\Delta t_o}{\Delta t_e} = \frac{\mathring{W}(t_e)}{\mathring{W}(t_o)}\,.
\end{equation}
In order to obtain the redshift, we finally need to calculate the ratio of the proper time intervals passing at the source and the observer. Since we assume that both the source and the observer are at rest with respect to the cosmological background, and hence obey \(w = 0\), the ratio of coordinate time \(t\) and proper time \(\tau\) is given by
\begin{equation}
\frac{dt}{d\tau} = |\tilde{L}(t, 0)|^{-\frac{1}{h}}\,.
\end{equation}
Using the abbreviation \(\mathring{L}(t) = \tilde{L}(t, 0)\) we thus find the redshift
\begin{equation}\label{eqn:redshift}
1 + z = \frac{\Delta\tau_o}{\Delta\tau_e} = \left(\frac{|\mathring{L}(t_o)|}{|\mathring{L}(t_e)|}\right)^{\frac{1}{h}}\frac{\Delta t_o}{\Delta t_e} = \left(\frac{|\mathring{L}(t_o)|}{|\mathring{L}(t_e)|}\right)^{\frac{1}{h}}\frac{\mathring{W}(t_e)}{\mathring{W}(t_o)} = \frac{\mathring{W}_L(t_e)}{\mathring{W}_L(t_o)}\,,
\end{equation}
where we have defined \(\mathring{W}_L(t) = \mathring{W}(t)|\mathring{L}(t)|^{-\frac{1}{h}}\). Note that we can always choose the coordinate time \(t\) such that it becomes identical to the proper time of an observer at rest with respect to the cosmological background, such that \(\mathring{L}(t) \equiv 1\). However, in section~\ref{sec:examples} we will encounter also examples for which a different choice of the coordinate time is more convenient, and so we will not impose a particular parametrization here. Note further that in Finsler spacetimes there may be more than one light cone, in which case there will be multiple solutions for \(\mathring{W}(t)\). This situation implies the existence of multiple types of light, and which would, in general, undergo different redshifts.

%------------------------------------------------------------------------------------------%
\subsection{Magnitude of a light source}\label{ssec:magnitude}
For convenience we assume in this section that the light source is located at the origin \(r_e = 0\), so that at the time of the observation, the light pulse forms a sphere of coordinate radius \(r_o\). We further assume that at the time \(t_e\) of the emission of radiation the source has a luminosity (radiation power) \(\mathfrak{L}\). The total radiation power flowing through this sphere as measured by an observer at \(r_o\) is influenced by the cosmological redshift in two ways: both the rate of photons and the frequency (and hence also the energy) of each photon, both measured using the respective proper times of the source and the observer, are reduced by a factor \(1 + z\), so that the observed power is
\begin{equation}\label{eqn:radpower}
\mathfrak{P} = \frac{\mathfrak{L}}{(1 + z)^2}\,.
\end{equation}
In order to calculate the magnitude of the light signal, we need the area of the illuminated sphere as measured in the rest frame of the observer given by \(y^t > 0,w = 0\), since this is the frame in which he also measures the detector area.

These areas are determined by the area measure induced from the Finsler metric via the determinant of its pullback to the surface of interest. Thus a well defined area measure requires a well defined second derivative of the Finsler geometry function $L$ at the position of the observer at rest. This can only be achieved if \(\partial_wL(t, y, w) = 0\) at \(y > 0, w = 0\), which implies in particular that \(\tilde{L}_w(t, 0) = 0\). The form of this condition arises from the fact that our coordinates have a coordinate singularity at \(w = 0\), and a geometry function \(L\) with \(\partial_wL(t, y, w) \neq 0\) at this point would possess a cusp. Taking this condition into account, the Finsler metric becomes
\begin{equation}\label{eqn:fmetric}
g^F_{ab}\,dx^a \otimes dx^b = \tilde{L}^{\frac{2}{h}}\,dt \otimes dt + \frac{1}{h}\tilde{L}^{\frac{2}{h} - 1}\tilde{L}_{ww}\left[\frac{dr \otimes dr}{1 - kr^2} + r^2\left(d\theta \otimes d\theta + \sin^2\theta\,d\phi \otimes d\phi\right)\right]\,.
\end{equation}
Note that we need \(\tilde{L}^{\frac{2}{h} - 1}\tilde{L}_{ww} < 0\) in order to have a metric with Lorentzian signature. In particular, we find the area of the sphere with coordinate radius \(r_o\) around the origin to be
\begin{equation}\label{eqn:radarea}
A = \frac{4\pi r_o^2}{h}\left|\tilde{L}^{\frac{2}{h} - 1}\tilde{L}_{ww}\right|\,.
\end{equation}
The radiation flux is thus given by
\begin{equation}\label{eqn:radflux}
\mathfrak{S} = \frac{\mathfrak{P}}{A} = \frac{h\mathfrak{L}}{4\pi r_o^2(1 + z)^2\left|\tilde{L}^{\frac{2}{h} - 1}\tilde{L}_{ww}\right|}\,.
\end{equation}
The magnitude is hence given by
\begin{equation}\label{eqn:magnitude}
m = -\frac{5}{2}\log_{10}\mathfrak{S} + \text{const.} = 5\log_{10}[r_o(1 + z)] + \frac{5}{2}\log_{10}\left|\tilde{L}^{\frac{2}{h} - 1}\tilde{L}_{ww}\right| - \frac{5}{2}\log_{10}\mathfrak{L} + \text{const.}
\end{equation}

%------------------------------------------------------------------------------------------%
\subsection{Relating magnitude and redshift}\label{ssec:magred}
We finally need to express the magnitude as a function of the redshift for a fixed observation time \(t_o\). For this purpose it is useful to first express both magnitude and redshift as functions of the emission time and then to take the inverse of the latter and substitute it into the former. We start with the redshift, which can be written as
\begin{equation}
z(t_e) = \frac{\mathring{W}_L(t_e)}{\mathring{W}_L(t_o)} - 1\,.
\end{equation}
The inverse of this relation is given by
\begin{equation}\label{eqn:timeredshift}
t_e(z) = \mathring{W}_L^{-1}[(1 + z)\mathring{W}_L(t_o)]\,,
\end{equation}
where we require \(\mathring{W}_L(t)\) to be invertible in the interval between \(t_e\) and \(t_o\). For the magnitude we can ignore the term involving \(\tilde{L}^{\frac{2}{h} - 1}\tilde{L}_{ww}\), since it is evaluated at the fixed observation time \(t_o\) and \(w = 0\), so that it can be absorbed into the additive constant. Besides the redshift we therefore only need the radius \(r_o\) of the sphere around the radiation source which is spanned by the signal at the observation time. This can be obtained from the integral
\begin{equation}
D(r_o) = \int_{0}^{r_o}\frac{dr}{\sqrt{1 - kr^2}} = \int_{t_e}^{t_o}\mathring{W}(t)\,dt\,,
\end{equation}
where \(D\) is the inverse of the function \(\Sigma\) defined by
\begin{equation}
\Sigma(x) = \sum_{i = 0}^{\infty}(-k)^i\frac{x^{2i + 1}}{(2i + 1)!} = \begin{cases}
\sin x & k = 1\,,\\
x & k = 0\,,\\
\sinh x & k = -1\,.
\end{cases}
\end{equation}
Thus, we have
\begin{equation}
r_o(t_e) = \Sigma\left(\int_{t_e}^{t_o}\mathring{W}(t)\,dt\right)\,.
\end{equation}
This finally yields the magnitude
\begin{equation}
m(t_e) = 5\log_{10}\left[\Sigma\left(\int_{t_e}^{t_o}\mathring{W}(t)\,dt\right)\frac{\mathring{W}_L(t_e)}{\mathring{W}_L(t_o)}\right] - \frac{5}{2}\log_{10}\mathfrak{L} + \text{const.}
\end{equation}
We see that both \(z(t_e)\) and \(m(t_e)\), and hence also \(m(z)\) are fully determined by the functions \(\mathring{W}\), describing light propagation on the cosmological Finsler background, and \(\mathring{W}_L\), describing the redshift. In order to determine \(m(z)\) in the recent past, i.e., for small \(z\), it is helpful to develop these functions in a Taylor series around the observation time \(t_o\) in the form
\begin{equation}
\mathring{W}(t) = \sum_{i = 0}^{\infty}\left.\frac{d^i\mathring{W}}{dt^i}\right|_{t = t_o}\frac{(t - t_o)^i}{i!} = \sum_{i = 0}^{\infty}\mathring{W}_i\frac{(t - t_o)^i}{i!}\,,
\end{equation}
and analogously for \(\mathring{W}_L\). For the redshift we then find the series expansion
\begin{equation}
z(t_e) = \frac{1}{\mathring{W}_{L0}}\left[\mathring{W}_{L1}(t_e - t_o) + \frac{1}{2}\mathring{W}_{L2}(t_e - t_o)^2 + \frac{1}{6}\mathring{W}_{L3}(t_e - t_o)^3 + \mathcal{O}((t_e - t_o)^4)\right]
\end{equation}
and its inverse
\begin{equation}
t_e(z) = t_o + \frac{\mathring{W}_{L0}}{\mathring{W}_{L1}}z - \frac{\mathring{W}_{L0}^2\mathring{W}_{L2}}{2\mathring{W}_{L1}^3}z^2 - \frac{\mathring{W}_{L1}\mathring{W}_{L3} - 3\mathring{W}_{L2}^2}{6\mathring{W}_{L1}^5}\mathring{W}_{L0}^3z^3 + \mathcal{O}(z^4)\,.
\end{equation}
Here we now further demand that also the inverse \(\mathring{W}_L^{-1}\) of \(\mathring{W}_L\) appearing in the underlying equation~\eqref{eqn:timeredshift} can be developed into a Taylor series around the corresponding point \(\mathring{W}_{L0}\), which in particular implies \(\mathring{W}_{L1} \neq 0\), since otherwise \(\mathring{W}_L^{-1}\) would not be differentiable at this point. The radius \(r_0\) takes the form
\begin{equation}
r_0(t_e) = -\mathring{W}_0(t_e - t_o) - \frac{1}{2}\mathring{W}_1(t_e - t_o)^2 - \frac{1}{6}\left(\mathring{W}_2 - k\mathring{W}_0^3\right)(t_e - t_o)^3 + \mathcal{O}((t_e - t_o)^4)\,.
\end{equation}
This yields the magnitude
\begin{multline}
m(t_e) = \text{const.} - \frac{5}{2}\log_{10}\mathfrak{L} + 5\log_{10}(t_e - t_o) + \frac{5}{2\ln 10}\left(\frac{\mathring{W}_1}{\mathring{W}_0} + 2\frac{\mathring{W}_{L1}}{\mathring{W}_{L0}}\right)(t_e - t_o)\\
+ \frac{5}{24\ln 10}\left(4\frac{\mathring{W}_2}{\mathring{W}_0} + 12\frac{\mathring{W}_{L2}}{\mathring{W}_{L0}} - 3\frac{\mathring{W}_1^2}{\mathring{W}_0^2} - 12\frac{\mathring{W}_{L1}^2}{\mathring{W}_{L0}^2} - 4k\mathring{W}_0^2\right)(t_e - t_o)^2 + \mathcal{O}((t_e - t_o)^3)\,,
\end{multline}
where all constant terms, including constant prefactors appearing inside logarithms, have been absorbed into the term called ``const''.
Finally, we find the magnitude-redshift relation
\begin{equation}\label{eqn:magred}
m(z) = 5\log_{10}z + \frac{5}{2\ln 10}\left(2 + \frac{\mathring{W}_1\mathring{W}_{L0}}{\mathring{W}_0\mathring{W}_{L1}} - \frac{\mathring{W}_{L0}\mathring{W}_{L2}}{\mathring{W}_{L1}^2}\right)z + \mathcal{O}(z^2) - \frac{5}{2}\log_{10}\mathfrak{L} + \text{const.}
\end{equation}
Comparing the coefficient in brackets with the conventionally used expression \(1 - q\) in terms of the deceleration parameter \(q\) finally yields
\begin{equation}\label{eqn:decpar}
q = \frac{\mathring{W}_{L0}\mathring{W}_{L2}}{\mathring{W}_{L1}^2} - \frac{\mathring{W}_1\mathring{W}_{L0}}{\mathring{W}_0\mathring{W}_{L1}} - 1\,.
\end{equation}
We see that the value of the deceleration parameter at the observation time is fully determined by the first three coefficients of the Taylor expansion of the functions \(\mathring{W}\) and \(\mathring{W}_L\). This result holds for any cosmologically symmetric Finsler spacetime with a well defined null structure and \(\mathring{W}_0 \neq 0, \mathring{W}_{L1} \neq 0\), so that the procedure above can be applied.

%------------------------------------------------------------------------------------------%
\subsection{Invariance under time reparametrization}\label{ssec:diffinv}
Since we have used an arbitrary parametrization for the cosmological time \(t\) in our derivation above, we finally discuss the invariance of the result under strictly monotonous reparametrizations \(t \to t'(t)\) of the time coordinate, while leaving the spatial coordinates \(r, \theta, \phi\) unchanged. Under this reparametrization the cosmological coordinates transform to \(y' = y\partial_tt'\) and \(w' = w\), where $\partial_tt'>0$. The geometry function, which is a scalar function on the tangent bundle, then transforms as
\begin{equation}
L'(t'(t), y'(t, y), w'(w)) = L'(t', y\partial_tt', w) = L(t, y, w)\,.
\end{equation}
For \(w = 0\) we find in particular
\begin{equation}
\mathring{L}'(t'(t)) = L'(t'(t), 1, 0) = L(t, (\partial_tt')^{-1}, 0) = L(t, 1, 0)(\partial_tt')^{-h} = \mathring{L}(t)(\partial_tt')^{-h}\,.
\end{equation}
Similarly, we can determine \(\mathring{W}'\) from
\begin{equation}
0 = L'(t'(t), 1, \mathring{W}'(t'(t))) = L(t, (\partial_tt')^{-1}, \mathring{W}'(t'(t))) = L(t, 1, \mathring{W}'(t'(t))\partial_tt')(\partial_tt')^{-h}\,,
\end{equation}
so that \(\mathring{W}'(t'(t)) = \mathring{W}(t)(\partial_tt')^{-1}\). In the following we will drop the arguments whenever they are clear from the context, and simply write \(\mathring{W} = \mathring{W}'\partial_tt'\) and \(\mathring{L} = \mathring{L}'(\partial_tt')^h\). Note in particular that
\begin{equation}
\mathring{W}_L' = \frac{\mathring{W}'}{|\mathring{L}'|^\frac{1}{h}} = \frac{\mathring{W}}{|\mathring{L}|^\frac{1}{h}} = \mathring{W}_L\,,
\end{equation}
which proves that the redshift \(z\) in~\eqref{eqn:redshift} is invariant, as necessary for an observable quantity relating proper time intervals. Also the luminosity \(\mathfrak{L}\) of the source is invariant, as it is defined using the proper time of the source. Equation~\eqref{eqn:radpower} then implies that also the radiation power \(\mathfrak{P}\) at the observation time is invariant. The Finsler metric~\eqref{eqn:fmetric} is invariant, which can most easily be seen from the transformations
\begin{equation}
dt = dt'(\partial_tt')^{-1}\,, \quad
\tilde{L}(t, 0) = \tilde{L}'(t', 0)(\partial_tt')^h\,, \quad
\tilde{L}_{ww}(t, 0) = \tilde{L}'_{w'w'}(t', 0)(\partial_tt')^{h - 2}\,.
\end{equation}
Hence, also the area~\eqref{eqn:radarea}, the flux~\eqref{eqn:radflux} and finally the magnitude~\eqref{eqn:magnitude} are invariant under reparametrization of the time coordinate, as expected. We thus conclude that also the magnitude-redshift relation \(m(z)\), its series expansion~\eqref{eqn:magred} and finally the deceleration parameter~\eqref{eqn:decpar} are invariant. To see this directly, one derives the transformation rules
\begin{gather}
\mathring{W}_0 = \mathring{W}_0'\partial_tt'|_{t_o}\,, \quad
\mathring{W}_1 = \mathring{W}_1'(\partial_tt'|_{t_o})^2 + \mathring{W}_0'\partial_t^2t'|_{t_o}\,, \quad
\mathring{W}_{L0} = \mathring{W}_{L0}'\,,\nonumber\\
\mathring{W}_{L1} = \mathring{W}_{L1}'\partial_tt'|_{t_o}\,, \quad
\mathring{W}_{L2} = \mathring{W}_{L2}'(\partial_tt'|_{t_o})^2 + \mathring{W}_{L1}'\partial_t^2t'|_{t_o}\,.
\end{gather}
Inserting these relations into the expression~\eqref{eqn:decpar} for the deceleration parameter then yields
\begin{equation}
q = \frac{\mathring{W}_{L0}\mathring{W}_{L2}}{\mathring{W}_{L1}^2} - \frac{\mathring{W}_1\mathring{W}_{L0}}{\mathring{W}_0\mathring{W}_{L1}} - 1 = \frac{\mathring{W}_{L0}'\mathring{W}_{L2}'}{\mathring{W}_{L1}'^2} - \frac{\mathring{W}_1'\mathring{W}_{L0}'}{\mathring{W}_0'\mathring{W}_{L1}'} - 1 = q'\,.
\end{equation}
This finally proves that also the deceleration parameter is invariant under a reparametrization of the time coordinate. We can thus apply our result to particular examples of Finsler spacetimes, and choose any suitable parametrization for the cosmological time. This will be done in the following section.

%%%%%%%%%%%%%%%%%%%%%%%%%%%%%%%%%%%%%%%%%%%%%%%%%%%%%%%%%%%%%%%%%%%%%%%%%%%%%%%%%%
\section{Examples}\label{sec:examples}
We now apply our calculation of the magnitude-redshift relation and the deceleration parameter of a general cosmological Finsler spacetime from the previous section to particular examples. As model independent example we consider a general first order Finsler perturbation of metric FLRW spacetime geometry, while afterwards we study particular non-perturbative Finsler spacetime models. The latter are split into two classes, those which have a light cone structure identical to metric FLRW geometry, and those which have a Finslerian light cone structure.

For all examples we calculate $\mathring{W}$ by solving equation~\eqref{eqn:geodredshift} and then $\mathring{W}_L$ is given by $\mathring{W}_L = \mathring{W} |\mathring{L}|^{-\frac{1}{h}}$. Together they determine the magnitude-redshift relation and the deceleration parameter according to equation \eqref{eqn:decpar}. For convenience we display the examples in their explicit coordinate form in the adapted coordinates $T$ and $W$ defined in equation~\eqref{eq:adapted}, and add a comment how they are defined covariantly in terms of tensor fields on spacetime.

%------------------------------------------------------------------------------------------%
\subsection{First order Finsler perturbations of FLRW geometry}\label{ssec:experturb}
As mentioned in the introduction of this article, when Finsler spacetime geometries are analyzed one often uses specific models, due to the large variety of possible Finsler spacetime geometries. In addition to our analysis of specific Finsler models in the previous and in the next subsection we consider a general first order Finslerian perturbation of FLRW geometry which takes the form
\begin{align}
	\tilde{L}(T, W) = (-1 + a(T)^2W^2)^\frac{h}{2} + \epsilon\tilde{G}(T,W)\,,
\end{align}
where the function $\tilde{G}(T,W)$ is generated from a function $G(t,y,w)$ which is $h$-homogeneous and reversible with respect to its last two arguments by setting
\begin{align}
	\tilde{G}(T,W) = \frac{1}{y^h}G(t,y,w) = G(T,1,W)\,,
\end{align}
in the same way as \(\tilde{L}\) is defined through \(L\). For $\epsilon = 0$ the geometry derived from $\tilde{L}(T, W)$ is the usual metric FLRW geometry for any choice of $h$, see \cite{Pfeifer:2011xi}. However, to obtain well-defined Finsler spacetimes we need to require that $h = 2n$ for some integer $n\geq 1$. Observe that if $n$ is itself even, then $\tilde L$ is positive, while for $n$ being odd $\tilde L$ is negative along the future pointing timelike curves. Both cases are of phenomenological interest. For $n=1$ the example is a Finsler perturbation of FLRW geometry, while for $n=2$ the example includes first order bi-metric geometries like the geometry which describes the propagation of light in an uniaxial crystal \cite{Pfeifer:2016har}.

Now determining \(\mathring{W}\) from the light cone condition $\tilde L(T,\mathring{W}) = 0$ requires solving
\begin{align}
	\mathring W^2 = \frac{1 + (-\epsilon\tilde{G}(T,\mathring{W}))^{\frac{1}{n}}}{a(T)^2}\,.
\end{align}
Since we are interested in the effect of the Finsler modification $\tilde{G}$ on the observables in FLRW geometry we make the Ansatz $\mathring{W} = \frac{1}{a(T)} + \mathring{W}_G$. We find that to the first perturbation order in $\mathring{W}_G$ the equation is solved~by
\begin{equation}\label{eqn:wperturb}
\mathring{W}_G = \left[2a(T)\left(-\epsilon\tilde{G}\left(T,\tfrac{1}{a(T)}\right)\right)^{-\frac{1}{n}} - \frac{2\partial_W\tilde{G}\left(T,\tfrac{1}{a(T)}\right)}{h\tilde{G}\left(T,\tfrac{1}{a(T)}\right)}\right]^{-1}
= \frac{1}{2a(T)}\left[-\epsilon\tilde{G}\left(T,\tfrac{1}{a(T)}\right)\right]^{\frac{1}{n}} + \mathcal{O}\left(\epsilon^{\frac{2}{n}}\right)\,.
\end{equation}
Thus the normalized $\mathring{W}_L$ becomes to leading order in $\epsilon$
\begin{align}
	\mathring{W}_L = \mathring{W} |\mathring{L}|^{-\frac{1}{2n}}
	&= \mathring{W}\left[1 - (-1)^n \frac{\epsilon}{2n} \tilde{G}(T,0) + \mathcal{O} (\epsilon^{2})\right] \approx \mathring{W}\,,
\end{align}
since the leading order in $\epsilon$ in the expression~\eqref{eqn:wperturb} contributes with $\epsilon^{\frac{1}{n}}$, which, for small $\epsilon$ has a larger effect than the order $\epsilon$ term in $\mathring{W}(-1)^n \frac{\epsilon}{2n} \tilde{G}(T,0)$. The only case in which this is not true is for $n=1$, where we find
\begin{align}
	\mathring{W}_L
	&\approx \mathring{W} + \frac{\epsilon}{2 a(T)}\tilde G(T,0) \,,
\end{align}
since here the order $\epsilon$ contribution from $- \frac{\epsilon}{2} \mathring{W} \tilde{G}(T,0)$ is of same order as the relevant order in $ \mathring{W}$ and so can not be neglected.

From these expressions we can derive the ingredients to calculate the deceleration parameter $q$ according to equation \eqref{eqn:decpar}, which simplifies for all models with $n\neq 1$ to
\begin{align}
	q|_{n\neq1} = \frac{\mathring{W}_{0}\mathring{W}_{2}}{\mathring{W}_{1}^2} - 2.
\end{align}
It is convenient to introduce a series expansion of the form
\begin{equation}\label{eqn:qseries}
a(T) = a_0\left[1 + H_0(T - T_o) - \frac{1}{2}H_0^2q_0(T - T_o)^2\right] + \mathcal{O}\left((T - T_o)^3\right)
\end{equation}
around the current time \(T_o\) in terms of the current time Hubble parameter \(H_0\) and deceleration parameter \(q_0\) for the scale factor \(a(T)\). In terms of these expansion parameters we find
\begin{multline}
	q|_{n\neq1} = q_0 + \frac{\epsilon^{\frac{1}{n}}}{2n^2}\frac{1}{a_0^2 H_0^2} (-\tilde{G}_{00} )^{\frac{1}{n} - 2}\bigg\{(1-n) (H_0 \tilde{G}_{01} - a_0 \tilde{G}_{10})^2 + n a_0^2 \tilde{G}_{00} \tilde{G}_{20}\\
	+ n \tilde{G}_{00} \left[ H_0^2 ( \tilde{G}_{02} - a_0 q_0 \tilde{G}_{01} ) + 2 a_0 H_0 ( a_0 (q_0+1) \tilde{G}_{10} - \tilde{G}_{11}) \right] \bigg\} + \mathcal{O}\left(\epsilon^{\frac{2}{n}}\right)\,,
\end{multline}
where, for the coefficients of the Taylor expansion of \(\tilde{G}\), we use the shorthand notation
\begin{equation}
\tilde{G}_{ij} = \frac{\partial^{i+j}\tilde{G}}{\partial T^i \partial W^j}\left(T_o, \tfrac{1}{a(T_o)}\right)\,.
\end{equation}
In the case $n=1$ we must employ the general formula
\begin{align}
	q|_{n=1} = \frac{\mathring{W}_{L0}\mathring{W}_{L2}}{\mathring{W}_{L1}^2} - \frac{\mathring{W}_{1}\mathring{W}_{L0}}{\mathring{W}_{0}\mathring{W}_{L1}} - 1\,,
\end{align}
which, using the notation introduced above, yields
\begin{multline}
	q|_{n=1} = q_0 - \frac{\epsilon}{2}\frac{1}{ a_0^2 H_0^2}\bigg\{ a_0^2 \tilde{G}_{20} + H_0^2 ( \tilde{G}_{02} - a_0 q_0 \tilde{G}_{01} ) \\
	+ 2 a_0 H_0 [ a_0 (q_0+1) \tilde{G}_{10} - \tilde{G}_{11}]
	- a_0^2 H_0 (1 + 2 q_0) \mathring{G}_{10} - a_0^2 \mathring{G}_{20}\bigg\} + \mathcal{O}\left(\epsilon^{2}\right)\,,
\end{multline}
where
\begin{equation}
\mathring{G}_{ij} = \frac{\partial^{i+j}\tilde{G}}{\partial T^i \partial W^j}\left(T_o, 0\right)\,.
\end{equation}
Now the deceleration parameter is a direct observable quantity for which we derived the influence of a Finslerian spacetime geometry to leading order. This yields the possibility to compare the modification of $q$ with the experimental data and to identify viable Finsler perturbations which are in agreement with observations. This result is one further step in the systematic phenomenological analysis of Finsler perturbations of the Lorentzian metric geometry of spacetime.

%------------------------------------------------------------------------------------------%
\subsection{FLRW metric null structure}\label{ssec:exmetric}
One way to construct Finsler spacetime geometries is to multiply the usual metric length measure by another function on the tangent bundle. Common examples in the literature are
\begin{itemize}
	\item
	The FLRW length measure itself
	\begin{equation}
	\tilde{L}(T, W) = -1 + a(T)^2W^2
	\end{equation}
	which is a metric Finsler spacetime $L = g_{ab}(x)y^ay^b$. In this example $|\mathring{L}| = \tilde L(T,0) = 1$ and so $\mathring{W}_L = \mathring{W}$. Using the notation introduced when discussing the perturbative example, one easily finds for $\mathring{W}$ and the deceleration parameter
	\begin{align}
		\mathring{W} = \frac{1}{a(T)},\quad q = q_0\,.
	\end{align}
	\item
	Bogoslovsky's length measure \cite{Bogoslovsky}, respectively the length measure of very special relativity \cite{Cohen:2006ky}, which was recognized in Finsler geometry \cite{Gibbons:2007iu} and also analyzed in cosmological context earlier \cite{Kouretsis:2008ha}:
	\begin{equation}
	\tilde{L}(T, W) = \left(-1 + a(T)^2W^2\right)b(T)^2.
	\end{equation}
	This length measure is constructed from a product between a metric length measure and a one-form, $L = \big(g_{ab}(x)y^ay^b\big)\big(A_c(x)y^c\big)^2$. Its null structure is given by the union of the FLRW light cone and the set of vectors $X$ which are annihilated by the one form $A$. In the discussion of the Bogoslovsky's length it is usually assumed that light propagates only on the metric light cone, which is the FLRW light cone here.
	On this part of the null-structure the solution for $\mathring{W}$ is the same as in the FLRW case, however, since $\mathring{W}_L\neq \mathring{W}$,
	\begin{align}
		\mathring{W} = \frac{1}{a(T)}, \quad \mathring{W}_L = \frac{1}{a(T)}\frac{1}{\sqrt{|b(T)|}}
	\end{align}
	the deceleration parameter is different in this class of Finsler spacetimes. If we use the same series expansion~\eqref{eqn:qseries} for \(a\) and a Taylor expansion of the form
	\begin{equation}\label{eqn:bseries}
	b(T) = \sum_{i = 0}^{\infty}\left.\frac{d^ib}{dT^i}\right|_{T = T_o}\frac{(T - T_o)^i}{i!} = \sum_{i = 0}^{\infty}b_i\frac{(T - T_o)^i}{i!}
	\end{equation}
	for \(b\), we obtain
	\begin{equation}
		q = \frac{H_0^2 q_0 + \frac{1}{2}\left(\frac{b_1}{b_0}\right)^2 - H_0 \frac{1}{2}\frac{b_1}{b_0} - \frac{1}{2}\frac{b_2}{b_0}}
		{(H_0 + \frac{1}{2}\frac{b_1}{b_0})^2}
		= q_0 - \frac{H_0(1+2 q_0) B_1 + B_2}{2 H_0^2} + \mathcal{O}(B^2)\,.
	\end{equation}
	In the last expression we approximated the free function $b(T)$ by $b(T) = 1 + B(T)$ and linearized $q$ in $B(T)$, using a Taylor expansion of the same form as~\eqref{eqn:bseries} for \(b\). We needed to expand $b(T)$ around $1$ since this is the case in which the Bogoslovsky length measure is identical to the FLRW length measure. An expansion around $b(T)=0$ does not make sense since then the length measure would vanish.
	\item
	An exponential modification of FLRW geometry, inspired by the example discussed in \cite{Minguzzi:2014aua},
	\begin{equation}
	\tilde{L}(T, W) = \left(-1 + a(T)^2W^2\right)\left(1 + e^{-\frac{(b(T))^2}{|-1 + a(T)^2W^2|}} \right)\,,
	\end{equation}
	which is a product of the metric length measure and the exponential of a zero-homogeneous function on the tangent bundle,
	\begin{equation}
	L = g_{ab}(x)y^ay^b\left[1 + \exp\left(-\frac{[A_a(x)y^a]^2}{|g_{ab}(x)y^ay^b|}\right)\right]
	\end{equation}
	determined by a one-form $A$ and a metric $g$. Again $\mathring{W}$ is identical to the FLRW case
	\begin{equation}
		\mathring{W} = \frac{1}{a(T)}, \quad \mathring{W}_L = \frac{1}{a(T)}\frac{1}{\sqrt{(1+e^{-b(T)^2})}},
	\end{equation}
	and $q$ is different
	\begin{equation}
	\begin{split}
	q &= \frac{(1 + e^{b_0^2})^2H_0^2 q_0 + (1 + e^{b_0^2})(b_1^2 + H_0b_1b_0 + b_2b_0) - 2 e^{b_0^2} b_0^2 b_1^2}
	{[H_0 (1 + e^{b_0^2}) - b_1b_0 ]^2}\\
	&= q_0 + \frac{b_1^2 + b_0 [H_0 (1+2q_0) b_1 + b_2 ] }{2 H_0^2} + \mathcal{O}(b^4)\,.
	\end{split}
	\end{equation}
	In the last line we expanded $q$ into the dominating order in $b$ around $b=0$, which is the quadratic order for this exponential length element.
\end{itemize}

These examples nicely show how a Finslerian spacetime geometry changes the prediction of the magnitude redshift relation, even though the light cone is not altered in these geometries compared to FLRW spacetime. Interestingly one finds for both, that in case $b_1 = b_2 = 0$, i.e., the first and second derivative of the function $b$ at the observation time vanish, the deceleration parameter is identical to the one predicted by FLRW spacetime geometry, $q=q_0$.

For a more fundamental theoretical prediction the only missing ingredient are dynamical equations which determine $a(T)$ and $b(T)$, just like the Einstein equations determine the scale factor in general relativity. Finsler generalization of the Einstein equations have been developed, for example in~\cite{Pfeifer:2011xi} or~\cite{Hohmann:2013fca}, however solving them for these examples is still work in progress and could not be achieved so far.

%------------------------------------------------------------------------------------------%
\subsection{Finslerian null structure}\label{ssec:exnonmetric}
Finsler spacetime geometries which alter the null-structure of metric spacetime geometry more fundamentally have different solutions for $\mathring{W}$ from equation \eqref{eqn:geodredshift}, as we have just seen. Here we consider the further simple examples of non-metric geometry functions:
\begin{itemize}
	\item
	The most general fourth order polynomial geometry
	\begin{equation}
	\tilde{L}(T, W) = -1 + a(T)^2W^2 + b(T)^4W^4
	\end{equation}
	is a Finsler spacetime geometry based on a general fourth rank tensor on spacetime, ${L=G_{abcd}(x)y^ay^by^cy^d}$. In this example the calculation of the deceleration parameter becomes surprisingly simple since $\mathring{W} = \mathring{W}_L$ by the fact that $|\mathring L| = |\tilde L(T,0)| = 1$. Solving for $\mathring{W}$ yields
	\begin{align}
	\mathring{W}^2 = \frac{1}{2 b^4}\left(-a^2+\sqrt{a^4+4b^4}\right)\,.
	\end{align}
	We consider in particular the case \(b \ll 1\) of a small perturbation of a geometry function given by
	\begin{equation}
	L(t, y, w) = y^4\tilde{L}(t, w/y) = y^2\left(-y^2 + a(t)^2w^2\right)\,,
	\end{equation}
	which as a factor contains a FLRW metric geometry with scale factor \(a\). The deceleration parameter is given by
	\begin{equation}
	q = q_0 + 2\frac{b_0^2}{a_0^4H_0^2}\left\{H_0^2(q_0 - 3)b_0^2 - 3b_1^2 - b_0[2H_0(q_0 - 3)b_1 + b_2]\right\} + \mathcal{O}(b^8)\,,
	\end{equation}
	where the lowest perturbation order is given by \(\mathcal{O}(b^4)\).

	\item
	An alternative fourth order ansatz is given by
	\begin{equation}
	\tilde{L}(T, W) = -1 + 2a(T)^2W^2 - [a(T) - b(T)]^4W^4\,,
	\end{equation}
	which leads to the solution
	\begin{equation}
	\mathring{W} = \frac{1}{\sqrt{a^2 + \sqrt{b(2a - b)(2a^2 - 2ab + b^2)}}}\,.
	\end{equation}
	Again $\mathring{W} = \mathring{W}_L$ as above. For \(b \ll a\) this can be viewed as a perturbation of the geometry function
	\begin{equation}
	L(t, y, w) = -\left(-y^2 + a(t)^2w^2\right)^2\,,
	\end{equation}
	which is essentially the square of the FLRW metric geometry function. Using the series expansion~\eqref{eqn:qseries} and the Taylor expansion~\eqref{eqn:bseries} we then find the deceleration parameter
	\begin{equation}
	q = q_0 + \frac{H_0^2(2q_0 + 1)b_0^2 + b_1^2 - 2b_0[H_0(2q_0 + 1)b_1 + b_2]}{4H_0^2\sqrt{a_0b_0^3}} + \mathcal{O}(b)\,.
	\end{equation}
	Note that the lowest perturbation order is given by \(\mathcal{O}\left(\sqrt{b}\right)\).

	\item
	Randers geometry, which is also discussed in the literature \cite{Randers} and known in physics as point particle action for a charged particle in an external electric potential
	\begin{equation}
	\tilde{L}(T, W) = \frac{1}{(y^t)^2}\left(\sqrt{|g_{ab}(x)y^ay^b|} + A_a(x)y^a\right)^2 = \left( \sqrt{|-1 + a(T)^2W^2|} + b(T)\right)^2\,,
	\end{equation}
	where we require \(b(T) < 0\) in order to obtain directions for which \(\tilde{L}\) becomes zero. Just like the Bogoslovsky example it is built from a metric and a one-form, $L = \left(\sqrt{g_{ab}(x)y^ay^b} + A_a(x)y^a\right)^2$. Observe that the Randers length does not define a Finsler spacetime according to our definition of Finsler spacetimes given in section \ref{ssec:finslerst}: the geometry function is neither smooth on $TM\setminus\{0\}$ nor reversible. However, we can derive the deceleration parameter of a cosmological Randers geometry here with the methods developed throughout this article, since for the derivation we only need the weaker condition that $L$ is smooth on its non-trivial null-directions and on the observer at rest.

	For $\mathring{W}$ and $\mathring{W}_L$ we obtain
	\begin{align}
	\mathring{W}^2 =\frac{1 - b^2}{a^2}, \quad \mathring{W}_L = \frac{\sqrt{1-b^2}}{a}\frac{1}{(1 + b)}\,,
	\end{align}
	which yield once more by using equation \eqref{eqn:decpar} the deceleration parameter $q$
	\begin{equation}
	\begin{split}
	q &=
	\frac{H_0^2 q_0 (b_0^2 -1 )^2 + b_0^2 (H_0 b_1 + b_2) - H_0 b_1 - b_2 - 2 b_0 b_1^2 - b_0b_1[H_0(1-b_0^2)+b_1]}
	{[H_0(1-b_0^2)+b_1]^2}\\
	&= q_0 - \frac{H_0(1+2 q_0) b_1 + b_2}{H_0^2} + \mathcal{O}(b^2)\,.
	\end{split}
	\end{equation}
	In the last line we again expanded $q$ to first order in $b$ around $b=0$. Surprisingly to first relevant order in the Finslerian effect the $q$ derived from the Randers length element and from the Bogoslovsky length element are nearly the same. The difference lies in a factor of~$\frac{1}{2}$.
	As in the examples with the FLRW light cone structure $b_1=b_2=0$ implies $q=q_0$.
\end{itemize}

It is clear that the interplay between the free functions in the length measure determines the redshift and hence the deceleration parameter. Again a quantitative statement can only be made after obtaining explicit expressions for these functions by solving the Finslerian gravitational dynamics. For the models presented in this section a deviation from the deceleration parameter as predicted by the standard $\Lambda$CDM model derived in general relativity already appears in the functional form of $\mathring{W}$, in contrast to the examples of section~\ref{ssec:exmetric}, where deviations enter only through \(\mathring{W}_L\).

%%%%%%%%%%%%%%%%%%%%%%%%%%%%%%%%%%%%%%%%%%%%%%%%%%%%%%%
\section{Discussion}\label{sec:discussion}
In this article we have discussed the geodesic motion and the propagation of light in Finsler spacetimes with cosmological symmetry. We have derived the geodesic equation using coordinates adapted to the cosmological symmetry, as well as a complete set of constants of motion which can be used to characterize geodesics and to reconstruct the geodesic equation. We have then discussed light propagation and derived expressions for the magnitude and redshift of a light source, from which we have obtained the magnitude-redshift relation for a general cosmologically symmetric Finsler spacetime as the central result of our work. In particular, we have derived a formula for the deceleration parameter in terms of the Finsler null structure.

From this general result we have derived the magnitude-redshift relation and deceleration parameter for several examples of Finsler spacetimes. In particular, we have discussed FLRW metric spacetime and its general perturbation, Bogoslovsky and Randers length measures, as well as exponential and fourth order corrections to FLRW spacetime. We have seen that at the kinematic level the deceleration parameter of several models agrees with that of FLRW metric spacetime, provided that they share the same null structure. For other models, which we treated as perturbations of FLRW metric spacetime, we have obtained leading order corrections to the deceleration parameter.

Note that our treatment of Finsler spacetimes has so far remained purely kinematic, since we have not assumed any particular action functional or dynamics for the Finsler geometry. For a full treatment of Finsler cosmology, of course also dynamics must be taken into account. We leave this full treatment, based on Finsler gravity action functionals as given, e.g., in \cite{Pfeifer:2011xi,Hohmann:2013fca}, for further investigation.

Our results allow a confrontation of cosmologically symmetric Finsler spacetimes with the most recent observations of supernovae~\cite{Kowalski:2008ez,Amanullah:2010vv,Suzuki:2011hu}, for which magnitude and redshift have been determined. The most simple approach would be a comparison of the deceleration parameters we have obtained to the values determined by previous analyses of supernova data~\cite{John:2004vf,Gong:2006gs,Cunha:2008mt,Li:2010da,Giostri:2012ek,Mukherjee:2016trt}. However, it should be noted that these values, in general, depend on the assumption of a particular kinematic model for the expansion of the universe, and may even depend on the parametrization of the magnitude-redshift relation, which complicates the analysis and makes a thorough, model-dependent study inevitable~\cite{Virey:2005ih,Linder:2008pp,Cattoen:2008th}. We will not enter this discussion here, as this would exceed the scope of this article, and leave this topic for future research.

%------------------------------------------------------------------------------%
\begin{acknowledgments}
MH gratefully acknowledges the full financial support of the Estonian Research Council through the Startup Research Grant PUT790 and the European Regional Development Fund through the Center of Excellence TK133 ``The Dark Side of the Universe''. CP gratefully thanks the Center of Applied Space Technology and Microgravity (ZARM) at the University of Bremen for their kind hospitality and acknowledges partial support of the European Regional Development Fund through the Center of Excellence TK133 ``The Dark Side of the Universe''.
\end{acknowledgments}

%%%%%%%%%%%%%%%%%%%%%%%%%%%%%%%%%%%%%%%%%%%%%%%%%%%%%%%
\appendix

%%%%%%%%%%%%%%%%%%%%%%%%%%%%%%%%%%%%%%%%%%%%%%%%%%%%%%%
\section{Complete lifts of the symmetry generators}\label{app:CL}
In manifold induced coordinates the complete lifts of the generators~\eqref{eqn:rotgen} of rotations displayed in section~\ref{ssec:cosmosym} are
\begin{subequations}
\begin{align}
\rho^{C}_1&=\sin\phi\partial_\theta+\cot\theta\cos\phi\partial_\phi+y^\phi\cos\phi\bar\partial_\theta-\Big(y^\theta\frac{\cos\phi}{\sin^2\theta}+y^\phi\cot\theta\sin\phi\Big)\bar\partial_\phi\,,\\
\rho^{C}_2&=-\cos\phi\partial_\theta+\cot\theta\sin\phi\partial_\phi+y^\phi\sin\phi\bar\partial_\theta-\Big(y^\theta\frac{\sin\phi}{\sin^2\theta}-y^\phi\cot\theta\cos\phi\Big)\bar\partial_\phi\,,\\
\rho^{C}_3&=\partial_\phi\,,
\end{align}
\end{subequations}
while the complete lifts of the translation generators~\eqref{eqn:transgen} are
\begin{subequations}
\begin{align}
\tau^{C}_1&=\chi\sin\theta \cos\phi \partial_r+\frac{\chi}{r}\cos\theta\cos\phi\partial_\theta-\frac{\chi}{r}\frac{\sin\phi}{\sin\theta}\partial_\phi\nonumber\\
&{}+\left(y^r\chi'\sin\theta\cos\phi+y^\theta\xi\cos\theta\cos\phi-y^\phi\xi\sin\theta\sin\phi\right)\bar\partial_r\nonumber\\
&{}+\Big(y^r\big(\frac{\chi}{r}\big)'\cos\theta\cos\phi-y^\theta\frac{\chi}{r}\sin\theta\cos\phi-y^\phi\frac{\chi}{r}\cos\theta\sin\phi\Big)\bar\partial_\theta\\
&{}+\Big(-y^r\big(\frac{\chi}{r}\big)'\frac{\sin\phi}{\sin\theta}+y^\theta\frac{\chi}{r}\frac{\sin\phi}{\sin^2\theta}\cos\theta-y^\phi\frac{\chi}{r}
\frac{\cos\phi}{\sin\theta}\Big)\bar\partial_\phi\nonumber\,,
\end{align}
\begin{align}
\tau^{C}_2&=\chi\sin\theta \sin\phi \partial_r+\frac{\chi}{r}\cos\theta\sin\phi\partial_\theta+\frac{\chi}{r}\frac{\cos\phi}{\sin\theta}\partial_\phi\nonumber\\
&{}+\left(y^r\chi'\sin\theta\sin\phi+y^\theta\xi\cos\theta\sin\phi+y^\phi\xi\sin\theta\cos\phi\right)\bar\partial_r\nonumber\\
&{}+\Big(y^r\big(\frac{\chi}{r}\big)'\cos\theta\cos\phi-y^\theta\frac{\chi}{r}\sin\theta\sin\phi+y^\phi\frac{\chi}{r}\cos\theta\cos\phi\Big)\bar\partial_\theta\\
&{}+\Big(y^r\big(\frac{\chi}{r}\big)'\frac{\cos\phi}{\sin\theta}-y^\theta\frac{\chi}{r}\frac{\cos\phi}{\sin^2\theta}\cos\theta-y^\phi\frac{\chi}{r}\frac{\sin\phi}{\sin\theta}\Big)\bar\partial_\phi\nonumber\,,
\end{align}
\begin{align}
\tau^{C}_3 = -\chi\cos\theta \partial_r+\frac{\chi}{r}\sin\theta\partial_\theta-\Big(y^r\chi'\cos\theta-y^\theta\chi\sin\theta\Big)\bar\partial_r+\Big(y^r\big(\frac{\chi}{r}\big)'\sin\theta+y^\theta\frac{\chi}{r}\cos\theta\Big)\bar\partial_\theta\,,
\end{align}
\end{subequations}
where we used the abbreviation \(\chi = \sqrt{1 - kr^2}\) and primes denote derivatives with respect to \(r\).

%%%%%%%%%%%%%%%%%%%%%%%%%%%%%%%%%%%%%%%%%%%%%%%%%%%%%%%
\section{The geodesic spray}\label{app:geodspray}
We want to calculate the components of the geodesic spray $G^a$ discussed in section~\ref{ssec:geodspray} in manifold induced coordinates $(t,r,\theta,\phi,y^t,y^r,y^\theta,y^\phi)$ on $TM$
\begin{align}\label{eq:geodspray1}
	G^a = \frac{1}{2}g^{L ab}\big(y^m\partial_m\bar\partial_bL-\partial_bL\big) \,.
\end{align}
The derivatives with respect to the $r,\theta$ and $\phi$ coordinates, as well as their corresponding directions $y^r,y^\theta,y^\phi$ are
\begin{equation}
\partial_\alpha L = \partial_w L \frac{\partial_\alpha w^2}{2 w}\,, \quad
\bar \partial_\alpha L = \partial_w L \frac{\bar\partial_\alpha w^2}{2 w} = \partial_w L \frac{y_\alpha}{w}\,,
\end{equation}
where $y_\alpha = w_{\alpha\beta}y^\beta$ and $w_{\alpha\beta} = \frac{1}{2}\bar\partial_\alpha\bar\partial_\beta w^2$ is the spatial part of the FLRW metric. The indices $\alpha,\beta,...$ run over $1 \sim r$, $2 \sim \theta$ and $3 \sim \phi$. All higher derivatives can be decomposed in an analogue way. Expanding the components of the geodesic spray yields
\begin{multline}\label{eq:geodspray2}
G^a = \frac{1}{2}g^{L at}\left( y^t \partial_t \bar\partial_t L + \partial_w \bar\partial_t L \frac{y^\alpha\partial_\alpha w^2}{2w} - \partial_t L \right) + \frac{1}{2}\frac{g^{L a\alpha}y_\alpha}{w} \left( y^t \partial_t \partial_w L + \partial_w \partial_w L \frac{y^\beta\partial_\beta w^2}{2w}\right)\\
+ \frac{1}{2}g^{L a\alpha}\left(y^\beta\partial_\beta \bar\partial_\alpha w - \partial_\alpha w\right)\partial_w L\,.
\end{multline}
The missing ingredient to completely calculate the components of the geodesic spray is the inverse of the $L$ metric. Observe that the $L$ metric takes the following form
\begin{equation}
	g^L_{ab} = \frac{1}{2}\bar\partial_a\bar\partial_b L \sim
	\left(\begin{array}{cc}
		\frac{1}{2}\bar\partial_t\bar\partial_t L & \frac{1}{2}\bar\partial_t\partial_w L \frac{y_\alpha}{ w} \\
		\frac{1}{2} \bar\partial_t\partial_w L \frac{y_\beta}{w} & \frac{\partial_w L}{2w} w_{\alpha\beta}+ \frac{1}{2} (\partial_w\partial_w L - \frac{\partial_w L}{w})\frac{y_\alpha}{w}\frac{y_\beta}{w}
	\end{array}\right)
	=\left(\begin{array}{cc}
		A & X_\alpha \\
		X_\beta & h_{\alpha\beta}
	\end{array}\right)\,.
\end{equation}
The inverse of a matrix of this form can be expressed in terms of the inverse $h^{\alpha\beta}$ of $h_{\alpha\beta}$ and $X^\alpha = h^{\alpha\mu}X_\mu$
\begin{align}
	g^{Lab} =
	\left(\begin{array}{cc}
		\frac{1}{A - h^{-1}(X,X)} & -\frac{X^\alpha}{A - h^{-1}(X,X)} \\
		-\frac{X^\beta}{A - h^{-1}(X,X)} & h^{\alpha\beta} + \frac{X^\alpha X^\beta}{A - h^{-1}(X,X)}
	\end{array}\right)\,.
\end{align}
Identifying the abbreviations with their definitions in terms of derivatives acting on $L$ we obtain
\begin{subequations}
\begin{align}
A &= \frac{1}{2}\bar\partial_t\bar\partial_t L\,,\\
h^{\alpha\beta} &= 2\bigg(\frac{w}{\partial_w L}w^{\alpha\beta} - \frac{(\partial_w\partial_w L - \frac{\partial_w L}{w})}{w \partial_w L\ \partial_w\partial_w L}y^\alpha y^\beta\bigg)\,,\\
X^\alpha &= \frac{1}{2 w}\bar\partial_t\partial_w L\ y_\beta h^{\alpha\beta} = \frac{y^\alpha}{ w}\frac{\bar\partial_t\partial_w L}{\partial_w\partial_w L}\,,\\
h^{-1}(X,X) &= X^\alpha X_\alpha = \frac{1}{2} \frac{(\bar\partial_t\partial_w L)^2}{\partial_w\partial_w L}\,.
\end{align}
\end{subequations}
We now use these expression to further simplify the components of the geodesic spray \eqref{eq:geodspray2}
\begin{align}
	G^t &= \frac{1}{2}g^{L tt}\big( y^t \partial_t \bar\partial_t L + \partial_w \bar\partial_t L \frac{y^\alpha\partial_\alpha w^2}{2w} - \partial_t L \big) - \frac{1}{2}g^{L tt}\frac{\bar\partial_t\partial_w L}{\partial_w\partial_w L} \big( y^t \partial_t \partial_w L + \partial_w \partial_w L \frac{y^\beta\partial_\beta w^2}{2w}\big)\nonumber\\
	&= \frac{1}{2}g^{L tt} (y^t \partial_t \bar\partial_t L - \frac{\bar\partial_t\partial_w L}{\partial_w\partial_w L} y^t \partial_t \partial_w L - \partial_t L)\,.
\end{align}
Additionally we can use the homogeneity of $L$ which implies $y^t \bar\partial_t L = h L - w \partial_w L$ to finally find
\begin{equation}\label{eqn:Gt}
	G^t = \frac{(h-1)}{2}\frac{g^{L tt}}{\partial_w\partial_w L}(\partial_t L\ \partial_w\partial_w L - \partial_w L\ \partial_t\partial_w L)\\
	= (h-1)\frac{\partial_t L\ \partial_w\partial_w L - \partial_w L\ \partial_t\partial_w L}{\partial_w\partial_w L\ \bar\partial_t\bar\partial_t L - (\partial_w\bar\partial_t L)^2}\,.
\end{equation}
The spatial components of the geodesic spray are
\begin{align}\label{eq:Gspatial}
	G^\alpha &= -\frac{1}{2}g^{L tt}X^\alpha \big( y^t \partial_t \bar\partial_t L - \partial_t L - \frac{\bar\partial_t\partial_w L}{\partial_w\partial_w L} \big( y^t \partial_t \partial_w L \big)\big)\nonumber\\
	&+ \frac{y^\alpha}{w \partial_w\partial_wL} \big( y^t \partial_t \partial_w L + \partial_w \partial_w L \frac{y^\beta\partial_\beta w^2}{2w}\big)
	+ w w^{\alpha\lambda}\big(y^\beta\partial_\beta \frac{\bar\partial_\lambda w^2}{2w} - \frac{1}{2w}\partial_\lambda w^2\big)\nonumber\\
	&= \frac{y^\alpha}{w \partial_w\partial_wL} \big( y^t \partial_t \partial_w L + \partial_w \partial_w L \frac{y^\beta\partial_\beta w^2}{2w} -G^t\bar\partial_t\partial_w L \big)
	+ w w^{\alpha\lambda}\big(y^\beta\partial_\beta \frac{\bar\partial_\lambda w^2}{2w} - \frac{1}{2w}\partial_\lambda w^2\big)\nonumber\\
	&= \frac{y^\alpha}{w} \bigg( y^t \frac{\partial_t \partial_w L}{\partial_w\partial_wL} + \frac{y^\beta\partial_\beta w^2}{2w} -G^t\frac{\bar\partial_t\partial_w L}{\partial_w\partial_wL} \bigg)
	+ w w^{\alpha\lambda}\bigg(y^\beta\partial_\beta \frac{y_\lambda}{w} - \frac{1}{2w}\partial_\lambda w^2\bigg)\nonumber\\
	&= \frac{y^\alpha}{w} \bigg( y^t \frac{\partial_t \partial_w L}{\partial_w\partial_wL} + \frac{y^\beta\partial_\beta w^2}{2w} -G^t\frac{\bar\partial_t\partial_w L}{\partial_w\partial_wL} \bigg)
	+ w^{\alpha\lambda}\bigg( y^\beta\partial_\beta y_\lambda- \frac{y_\lambda}{ 2w^2}y^\beta\partial_\beta w^2 - \frac{1}{2}\partial_\lambda w^2\bigg)\nonumber\\
	&= \frac{y^\alpha}{w} \bigg( y^t \frac{\partial_t \partial_w L}{\partial_w\partial_wL} -G^t\frac{\bar\partial_t\partial_w L}{\partial_w\partial_wL} \bigg)
	+ \frac{1}{2} y^\beta y^\sigma w^{\alpha\lambda}\bigg( \partial_\beta w_{\lambda\sigma} + \partial_\sigma w_{\lambda\beta} - \partial_\lambda w_{\beta\sigma}\bigg)\,.
\end{align}

%%%%%%%%%%%%%%%%%%%%%%%%%%%%%%%%%%%%%%%%%%%%%%%%%%%%%%%
\section{Radial geodesics}\label{app:radialgeod}
Here we present the radial geodesics derived in section \ref{ssec:radialgeo} in manifold induced coordinates, in which the geodesic equations take the form
\begin{align}
\ddot x^a + G^a(x,\dot x) = 0\,.
\end{align}
Since $\theta, \phi, y^\theta$ and $y^\phi$ are fixed to be $(\frac{\pi}{2}, 0, 0, 0)$ we first check, with help of equation \eqref{eq:Gspatial}, that the corresponding geodesic equations are satisfied:
\begin{subequations}
\begin{align}
\ddot \theta + G^\theta &= 0 + 2 w w^{\theta\theta}\left[\dot r\partial_r\left(\frac{1}{w}r^2 \dot{\theta}\right) - \frac{1}{w}r^2 \sin\theta \cos\theta \dot{\phi}^2\right] = 0 \,,\\
\ddot \phi + G^\phi &= 0 + 2 w w^{\phi\phi} \dot r\partial_r\left(\frac{1}{w}r^2\sin\theta^2 \dot{\phi}\right) = 0 \,.
\end{align}
\end{subequations}
The remaining two geodesic equations are
\begin{equation}
\ddot t + G^t = 0,\qquad \ddot r + G^r = \ddot r + \sqrt{1-k r^2} \bigg( \dot t \frac{\partial_t \partial_w L}{\partial_w\partial_wL} -G^t\frac{\bar\partial_t\partial_w L}{\partial_w\partial_wL} \bigg)
+ \dot r ^2 \frac{kr}{1-kr^2}=0\,,
\end{equation}
where \(G^t\) is given in equation~\eqref{eqn:Gt} we used the fact that
\begin{equation}
\frac{1}{2}w^{rr}\partial_rw_{rr} = \frac{kr}{1 - kr^2}\,.
\end{equation}
In addition we have for radial geodesics
\begin{equation}
\dot{r} = y^r = w\sqrt{1 - kr^2}\,,
\end{equation}
which implies that
\begin{equation}
\ddot r = \dot w \sqrt{1-kr^2} - w^2 k r\,.
\end{equation}
Hence, the radial equation simplifies to
\begin{align}
\dot w + \bigg( \dot t \frac{\partial_t \partial_w L}{\partial_w\partial_wL} -G^t\frac{\bar\partial_t\partial_w L}{\partial_w\partial_wL} \bigg) =0\,.
\end{align}
This is consistent with the fact that for radial geodesics we have one remaining non-vanishing constant of motion \(C_1 = \partial_wL = y^{h - 1}\tilde{L}_w\), which implies
\begin{equation}
0 = \frac{d}{d\lambda}C_1 = \partial_t\partial_w L \dot t + \bar{\partial_t}\partial_w L \ddot t + \partial_w \partial_wL \dot w\,.
\end{equation}
Solving for \(\dot{w}\) then yields
\begin{equation}
\dot w = (-\partial_t\partial_w L \dot t + \bar{\partial_t}\partial_w L G^t )\frac{1}{\partial_w \partial_wL}\,.
\end{equation}
Employing the fact that $C_0 = L$ itself is constant along the geodesics allows to eliminate the $\dot t$ in terms of $t$ and $w$ from the equations. Thus to solve for radial geodesics on a homogeneous and isotropic Finsler spacetime it suffices to solve $\partial_w L (t, \dot t, w) = \text{const.}$ for $w(t)$. This expression then can be integrated to obtain $r(t)$.

%%%%%%%%%%%%%%%%%%%%%%%%%%%%%%%%%%%%%%%%%%%%%%%%%%%%%%%
\bibliographystyle{utphys}
\bibliography{FR}

\end{document}